\documentclass[12pt]{iopart}
\usepackage{iopams}  

\usepackage{graphicx}  
\usepackage{dcolumn}   
\usepackage{bm}        
\usepackage{xspace}
\usepackage{xcolor}
\usepackage{float}
\usepackage{subcaption}

\usepackage [T1]{fontenc}
\usepackage [utf8]{inputenc}
\usepackage[paperwidth=210mm,paperheight=297mm,centering,hmargin=2cm,vmargin=2.5cm]{geometry}
\usepackage{multirow}

\usepackage[square,numbers,sort&compress]{natbib}
\usepackage[colorlinks]{hyperref}
\newcommand{\text}{\mathrm}
\newcommand{\eqref}[1]{(\ref{#1})}
\newcommand{\be}{\begin{equation}}
	\newcommand{\ee}{\end{equation}}
\newcommand{\bea}{\begin{eqnarray}}
	\newcommand{\eea}{\end{eqnarray}}

\graphicspath{{./Figures/}{./}}

\begin{document}

\title{Lack of self-averaging of the critical internal energy in a weakly-disordered Baxter model}

\author{Ramgopal Agrawal$^1$, Victor Dotsenko$^{2}$, Maxym Dudka$^{3,4,5}$, Marco Picco$^6$, Enzo Marinari$^{1,7}$, and Gleb Oshanin$^{2}$}

\address{$^1$Dipartimento di Fisica, Sapienza Universit\'a di Roma, P.le A. Moro 2, I-00185, Roma, Italy}
\address{$^2$Laboratoire de Physique Th{\'e}orique de la Mati{\`e}re Condens{\'e}e (UMR CNRS 7600),\\ Sorbonne Universit{\'e}, 4 Place Jussieu, 75252 Paris, France}
\address{$^3$Yukhnovskii Institute for Condensed Matter Physics, National Acad. Sci. of Ukraine,\\ 1 Svientsitskii str., 79011 Lviv, Ukraine}
\address{$^4$ ${\mathbb L}^4$  Collaboration \& Doctoral College for the Statistical Physics of Complex Systems,\\ Leipzig-Lorraine-Lviv-Coventry, Europe}
\address{$^5$Lviv Polytechnic National University, 12 S. Bandera str., 79013, Lviv, Ukraine}
\address{$^6$Laboratoire de Physique Th{\'e}orique et Hautes Energies (UMR CNRS  7589),\\ Sorbonne Universit{\'e}, 4 Place Jussieu, 75252 Paris, France}
\address{$^7$Nanotech-CNR, UOS di Roma and INFN, Sezione di Roma, P.le A. Moro 2, I-00185, Roma, Italy}

\begin{abstract}
\noindent
We investigate the first two moments of the critical internal energy $E$ in a weakly disordered two-dimensional Baxter eight-vertex model as a function of the system size $L$, evaluated at the pseudo-critical point. Disorder is introduced via an equivalent representation of the pure eight-vertex model in terms of two ferromagnetic Ising models coupled by a four-spin interaction of strength $g_0$, where the Ising couplings consist of a uniform ferromagnetic part $J>0$ supplemented by weak Gaussian spatial disorder. In the critical regime, the model is formulated in terms of interacting Grassmann--Majorana spinor fields with quartic interactions and analyzed, for small positive $g_0$, using a combination of replica and renormalization-group methods. We also run extensive numerical simulations measuring the critical internal energy. Our results show that its relative variance increases with $L$ and approaches a finite constant as $L \to \infty$ for both $\pm g_0$. Hence, fluctuations remain relevant independently of the sign of $g_0$ (and thus of the specific-heat exponent), implying a lack of self-averaging of both the critical internal energy and the free energy. Consequently, reliable estimates of these quantities require averaging over many disorder realizations. In addition, we numerically confirm earlier predictions concerning the absence of self-averaging of the critical internal energy in the disordered Ising model.
\end{abstract}

\vspace{1pc}



\maketitle

\pagestyle{plain}
\tableofcontents

\section{Introduction and Background}
\label{I}

The behavior of disordered spin systems at criticality is often remarkably different from the one shown by their pure counterparts. Following early seminal works~\cite{Harris74,Harris_Lub74,Khmelnitskii75,Lubensky75,Grinstein_Luther76,kinzel,DD82,Newman_Riedel82,Birgeneau83}, many intriguing disorder-induced features have been revealed and explained via diverse theoretical approaches~\cite{rep1,perturbedCFT1,perturbedCFT2,IGLOI2005277} as well as via extensive numerical analyses~\cite{Picco_2006,PhysRevE.108.064131}. On the other hand, some important questions still remain unanswered. In particular, there is evidence that some thermodynamic properties lack self-averaging in a few known examples of disordered spin systems (see, e.g., Refs.~\cite{NonSelfAverage1,NonSelfAverage2,NonSelfAverage3,NonSelfAverage4,Dotsenko_Klumov2012,Dotsenko_Holovatch14}). Concurrently, there are counterexamples in which these properties are self-averaging, albeit few are known (see, e.g.,~\cite{2DRIM-IntEnergy}). This paper is an attempt to answer this question with both analytical and numerical evidences for the case of a weakly-disordered two-dimensional Baxter model.

Over the past years, a substantial progress has been achieved in understanding what is the actual impact of a  weak quenched disorder on the critical properties of systems which exhibit a continuous phase transition~\cite{Harris74,kinzel}. It was demonstrated on several particular examples, and also using some general arguments known by now as the \textit{Harris criterion}~\cite{Harris74,Harris_Lub74}, that if the specific heat critical exponent $\alpha_{p}$ of the corresponding pure system is negative, $\alpha_{p} < 0$, i.e., the specific heat stays finite at the critical temperature $T_{c}$, then the presence of a weak disorder does not alter the critical behavior. On the contrary, if $\alpha_{p} >0$, i.e.,  the specific heat diverges at $T_{c}$, then a new universal critical regime sets in. In this regard, the exactly-solvable two-dimensional Ising model~\cite{Onsager44} appears to be particularly interesting, because it corresponds to the border-line case with $\alpha_{p} = 0$. Here the specific heat $C_{p}(\tau)$ of the pure model is logarithmically divergent at the critical point, $C_{p}(\tau) \, \sim \, \ln\bigl(1/|\tau|\bigr)$ (where reduced-temperature $\tau = (T-T_{c})/T_{c}$). It was shown that for the Ising model with a small concentration of impure bonds the specific heat $C_{imp}(\tau)$ remains divergent, although in a weaker way: $C_{imp}(\tau) \, \sim \, \ln\ln\bigl(1/|\tau|\bigr)$ \cite{DD82,DD_rev83}. The  critical properties of the two-dimensional disordered Ising model were scrutinized through a series of analytical and numerical studies; see Refs.~\cite{Picco_2006,PhysRevE.108.064131,2DRIM-rev1,2DRIM-rev2} and references therein. Apart of that, an impact of a structural disorder with correlated impurities has been studied in detail in Refs.~\cite{correlated1,correlated2,correlated3}. Moreover, the effects of long-ranged power-law correlations in more complicated spin systems, e.g.,  the Baxter model, or the $N$-color Ashkin-Teller model, have also been studied in Ref.~\cite{fed}.

A salient feature of systems with quenched disorder, unrelated to the Harris criterion and to changes in critical behavior, is the possible lack of self-averaging (see definition below) of certain thermodynamic quantities in the vicinity of the critical point~\cite{NonSelfAverage1,NonSelfAverage2,NonSelfAverage3,NonSelfAverage4,Dotsenko_Klumov2012}. A standard line of thought, first advocated by Brout~\cite{brout}, is based on the argument that an extensive thermodynamic quantity $X$, ($X$ may stand for the internal energy $E$, magnetization, specific heat, or susceptibility), in a random system must be self-averaging due to the following circumstance: away from the critical point and at scales much larger than the correlation length $\xi_{\rm c}$, a given system may be considered as a macroscopic number $N$ of essentially independent regions of size $\xi_{\rm c}$. Then, any extensive quantity $X$ being a sum of a large number $N$ of independent random contributions, should be normally distributed around its mean value, and its dimensionless relative variance $R_X$ (see, e.g., \cite{NonSelfAverage1}),
\begin{equation}
	\label{relVar}
	R_X \equiv \frac{\overline{X^2} - \overline{X}^2}{\overline{X}^2}
	\; ,
\end{equation}
should vanish with the system's size $L$ as $R_X \sim 1/N \sim L^{-d}$, where $d$ is the spatial dimension and $\overline{(\cdots)}$ denotes the average over disorder realizations. A thermodynamic observable for which $R_X$ behaves as $R_X \sim L^{-d}$ is told to be \textit{strongly} self-averaging. In such systems, fluctuations around the mean value are progressively less important the larger $L$ is and it clearly suffices for any required measurement to have a single very large sample, which will represent the behavior in an ensemble of all possible samples.

However, Brout's argument~\cite{brout} cannot be applied  for a system of a large but finite size $L$ that is sufficiently close to the criticality, i.e., when the correlation length $\xi_{\rm c}(\tau) \sim L$ (this also defines the so called {\it pseudo-critical point} $\tau_{*}$). In such a situation, the regions can be no longer considered as independent and the validity of the self-averaging hypothesis becomes at least not evident; in consequence, the relation $R_X \sim L^{-d}$ may be violated. In fact, the renormalization group (RG) analysis developed in Ref.~\cite{NonSelfAverage2} suggested that the relative variance should rather approach a constant $L$-independent value as $L \to \infty$. Such a behavior has been indeed seen for several examples: for a family of random-bond Ashkin-Teller models with $\alpha_p > 0$~\cite{NonSelfAverage1}, for a site-diluted Ising model on a cubic lattice~\cite{NonSelfAverage3}, as well as for the $p$-state mean-field Potts glass~\cite{NonSelfAverage4}. A lack of self-averaging of the free energy at the pseudo-critical point of the Ising model in dimensions $d < 4$ has been analytically proven~\cite{Dotsenko_Holovatch14}. Moreover, using a combined RG and replica approach it was demonstrated that the critical internal energy $E$ of the two-dimensional disordered Ising model is also not self-averaging at the pseudo-critical point~\cite{2DRIM-IntEnergy}. On the other hand, this list of known examples is not that long and also there are counterexamples of systems and extensive thermodynamic observables which do exhibit a self-averaging with $R_X \sim L^{-d}$, or in a weaker form $R_X \sim L^{-\nu}$ with $0< \nu < d$ (see, e.g., \cite{NonSelfAverage3,2DRIM-IntEnergy}).

In this regard, a more systematic analysis of different models and therefore, a more representative nomenclature are highly desirable to draw more definite conclusions, which are important because a lack of self-averaging has quite far reaching consequences: unlike it happens for systems with a self-averaging, here any measurement performed on a single (no matter how large) sample will not be representative of the behavior in an ensemble, implying that the measurements must be performed on a large number of samples. This signifies that the only meaningful property which provides an exhaustive information about $X$ is its full probability density function $P(X)$~\cite{NonSelfAverage3}.

\begin{figure}[t!]
\centering
\includegraphics[width=0.8\textwidth]{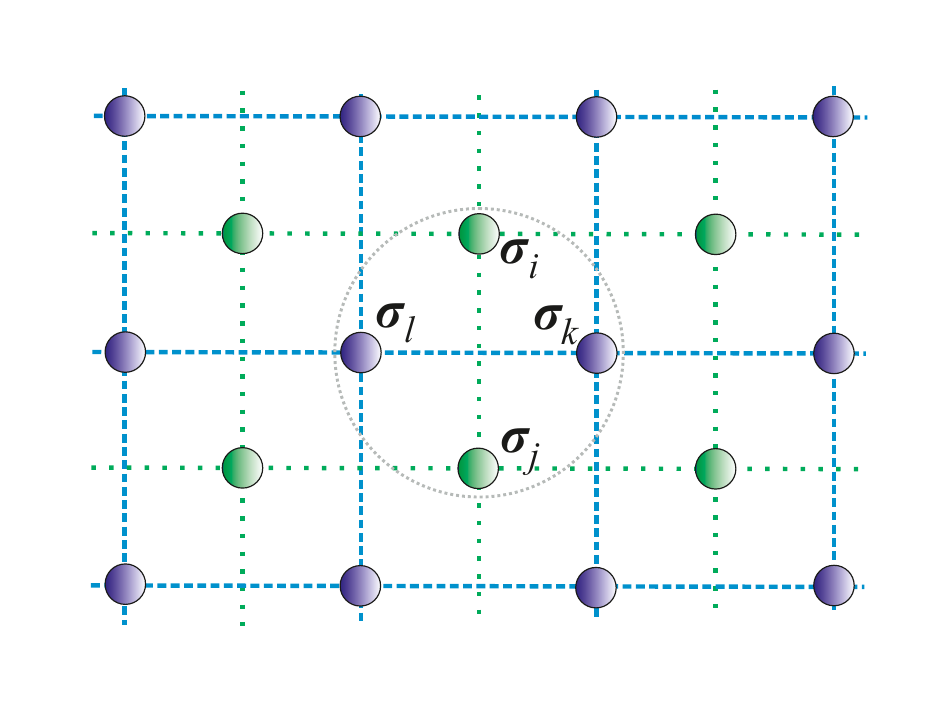}
	\caption{\textbf{Baxter's eight-vertex model:} The model is represented as two square-lattice Ising models coupled via four-spin interactions~\cite{Kadanoff_Wegner71,Wu71}. The first lattice (depicted by dotted lines) contains spins $\sigma^{(1)}$ (green) numbered by indices $i$ and $j$. The second (depicted by dashed lines) contains spins $\sigma^{(2)}$ (blue) numbered by indices $k$ and $l$. Additional four-spins interaction term involves the sum of products of values of the spins appearing within the dashed circle, its center being translated over all the crossings of the dashed and dotted lines.}
	\label{Fig:1}
\end{figure}

In this paper we inquire about a possible lack (or on contrary, a presence) of self-averaging of the critical internal energy $E$ in a weakly-disordered version of the Baxter's eight-vertex model~\cite{Baxter71}, at the above-defined pseudo-critical point $\tau_{*}$ in the limit when the system's size $L \to \infty$. To introduce disorder (which is not an evident issue for the eight-vertex model itself), we use the rigorous representation of the eight-vertex model in a form which is somewhat easier to grasp: it was shown in Ref.~\cite{Kadanoff_Wegner71,Wu71} that the eight-vertex model is equivalent to two Ising models coupled by four-spin interactions (see Fig.~\ref{Fig:1}). Using such an equivalence, the Hamiltonian of the model can be written as
\begin{eqnarray}
	\label{ham}
	H &=& -\sum_{<i,j>} J^{(1)}_{ij} \, \sigma^{(1)}_{i} \sigma^{(1)}_{j}
	-\sum_{<k,l>} J^{(2)}_{kl} \, \sigma^{(2)}_{k} \sigma^{(2)}_{l} \nonumber\\
	&-&g_{0} \sum_{<i,j,k,l>}  \sigma^{(1)}_{i} \sigma^{(1)}_{j} \sigma^{(2)}_{k} \sigma^{(2)}_{l}
	\; ,
\end{eqnarray}
where $J^{(1)}_{ij}$ and $J^{(2)}_{kl}$ define the interaction strengths between the nearest-neighboring (NN) Ising spins $\sigma^{(1)} = \pm 1$, and between the NN spins $\sigma^{(2)}  = \pm 1$ occupying the sites of the first (depicted by dotted lines in Fig.~\ref{Fig:1}) and of the second (depicted by dashed lines) square lattices, respectively. The term in the second line in Eq.~\eqref{ham} is crucial and introduces non-trivial {\it four-spin} interactions; namely, the interactions between two Ising spins in the first lattice and two Ising spins in the second one, as it is shown in Fig.~\ref{Fig:1}. The strength of these interactions is controlled by the coupling parameter $g_{0}$, which can be positive or negative (while the case $g_{0}=0$ corresponds to two independent Ising models). A (weak) disorder is introduced by assuming that the interaction strengths have the same non-fluctuating part $J > 0$ and an additional, position-dependent random Gaussian component with a small variance (see discussion below). Some results on an alternatively defined disordered Baxter's model with site impurities can be found in Ref.~\cite{swen}.

In the pure symmetrical case, the interaction strengths on the two Ising lattices are all equal to the same positive constant: $J^{(1)}_{ij} \, = \, J^{(2)}_{kl} \, = \, J > 0$.  A key feature here is that the critical exponents appear to be {\it non universal}, in the sense that they are continuously dependent on $g_{0}$, such that tuning the latter one may get negative or positive values of the specific heat exponent. More specifically, using the fermion representation of this model and the RG approach, it was shown~\cite{Luther_Peschel75,Luther76} that for $\vert g_0 \vert \ll 1$ the specific heat obeys $C_{p} \sim |\tau|^{-4g_{0}/\pi}$, which signifies that the corresponding critical exponent $\alpha_p$ can be positive or negative, depending on the sign of the coupling parameter. This important circumstance underlies our motivation for the choice of the model --- the latter provides a unified framework in which the impact of a positive or a negative value of $\alpha_p$ on a self-averaging characteristic can be probed within the same settings.

Our analytical approach extends the methodology developed previously in Ref.~\cite{2DRIM-IntEnergy} for the random $2d$ Ising model.  In the critical region, our model is re-formulated in terms of two interacting spinor fields with four-spin interactions. Using then a combination of the replica method and the RG technique, we derive the RG flow equations in the double limit
\begin{equation}
	\label{za}
	|g_0| \ll 1 \,,  \quad |g_0| \ln L \to \infty
	\,.
\end{equation}
which are subsequently solved for small positive values of $g_0$. In doing so, we show that the relative variance $R_E$ in Eq.~\eqref{relVar} approaches a constant value, meaning that (at least) for positive $g_0$ the critical  internal energy is not a self-averaging property.

We also numerically confirm the lack of self-averaging of the critical energy by simulating the weakly disordered Baxter model~\eqref{ham} on finite-size lattices, considering both positive and negative values of $g_0$. In addition, we validate the correctness of our analytical framework by providing clear numerical evidence for the disordered Ising model~\cite{2DRIM-IntEnergy}. We recall that in~\cite{2DRIM-IntEnergy} the critical internal energy was shown to exhibit a lack of self-averaging (using a combined setup of RG and replica trick methods). However, its numerical confirmation remained unresolved. The present paper provides a complete account, including both the extension of the methodology to the disordered Baxter model and the numerical verification of the claims for both the Baxter and Ising models.

The rest of the paper is organized as follows. In Sec.~\ref{fermion}, we introduce the disordered Baxter model and its continuum representation. In Sec.~\ref{replica}, the replica approach is discussed, and the replicated \textit{effective} Hamiltonian is derived. In Sec.~\ref{RG}, we obtain the simplified RG flow equations for the different couplings in the replicated Hamiltonian. In Sec.~\ref{moments}, the first two moments of the critical internal energy are calculated using the solutions of the RG flow equations for small positive $g_0$. In Sec.~\ref{numerical}, we present simulation results for the disordered Baxter model for both positive and negative value of parameter $g_0$. Finally, Sec.~\ref{conclusion} summarizes and discusses the results of this paper. Appendix~\ref{2DRIM} presents additional numerical results for the disordered Ising model.

\section{A weakly-disordered Baxter model and its continuous-space representation}\label{II}
\label{fermion}

\subsection{A weakly-disordered Baxter model}

The disordered Baxter model considered in this paper is defined by the Hamiltonian~\eqref{ham}, in which the interaction strengths of the two Ising models contain weak quenched randomness. Namely, we stipulate that $J^{(1)}_{ij} \, = \, J \, + \delta J^{(1)}_{ij}$ and $J^{(2)}_{kl} \, = \, J \, + \delta J^{(2)}_{kl}$, where $\delta J^{(1)}_{ij}$ and $\delta J^{(2)}_{kl}$ are mutually-independent zero-mean Gaussian random variables, such that
\begin{eqnarray}
	\nonumber
	&&\overline{ \delta J^{(1)}_{ij}\delta J^{(2)}_{kl} } \; = \;  0
	\; ,
	\\
	\label{gauss}
	\\
	\nonumber
	&&\overline{ \bigl(\delta J^{(1)}_{ij}\bigr)^{2} } \; = \; 
	\overline{ \bigl(\delta J^{(2)}_{kl}\bigr)^{2} } \; = \;  u_{0} \; .
\end{eqnarray}
Here and in the following, $\overline{(\cdots)}$ denotes the averaging over disorder, while the variance $u_{0} \ll J^{2}$ defines the strength of the disorder.

\subsection{Continuous-space representation}

As shown in Ref.~\cite{DD84}, in the vicinity of the critical temperature $T_{c}$ the scaling behavior of the pure Baxter model can be well described in terms of two real Grassmann-Majorana spinor fields $\psi({\bf r}) = \big(\psi_{1}({\bf r}), \, \psi_{2}({\bf r})\big)$ and $\chi({\bf r}) = \big(\chi_{1}({\bf r}), \, \chi_{2}({\bf r})\big)$, and the corresponding  Hamiltonian of the form:
\begin{eqnarray}
	\label{1}
	&&H_{0}[\psi,\chi; \tau] \; = \; \int d^{2}r \, \Big[
	-\frac{1}{2}\overline{\psi}({\bf r})\hat{\partial} \psi({\bf r}) \; -\frac{1}{2}\overline{\chi}({\bf r})\hat{\partial} \chi({\bf r}) \; \nonumber\\ &-& \; \frac{1}{2}\tau \, \overline{\psi}({\bf r})\psi({\bf r})
	- \; \frac{1}{2}\tau \, \overline{\chi}({\bf r})\chi({\bf r})+g_0(\overline{\psi}({\bf r})\psi({\bf r}))(\overline{\chi}({\bf r})\chi({\bf r}))
	\Big] \,,
\end{eqnarray}
where $\tau \propto (T-T_{c})/T_{c} \,  \ll \, 1$. In Eq.~\eqref{1} we use standard notations
\begin{equation}
	\label{2}
	\overline{\psi} \; \equiv \; \psi \, \hat{\sigma_{3}},  \quad \overline{\chi} \; \equiv \; \chi \, \hat{\sigma_{3}} \,, \quad
	\hat{\partial} \; = \; \hat{\sigma}_{1} \frac{\partial}{\partial x} \; + \;
	\hat{\sigma}_{2} \frac{\partial}{\partial y} \,,
\end{equation}

\begin{equation}
	\label{3}
	\hat{\sigma}_{1} =
	\left(
	\begin{array}{cc}
		0 & 1 \\
		1 & 0
	\end{array}
	\right), \qquad
	\hat{\sigma}_{2} =
	\left(
	\begin{array}{cc}
		0 & -i \\
		i & 0
	\end{array}
	\right), \qquad
	\hat{\sigma}_{3} =
	\hat{\sigma}_{1}\hat{\sigma}_{2} =
	\left(
	\begin{array}{cc}
		i & 0 \\
		0 & -i
	\end{array}
	\right).
\end{equation}
A weak quenched disorder can be introduced into Eq.~\eqref{1} by adding quenched spatial fluctuations to the mass parameter $\tau$, independently for the spinor fields $\psi$ and for $\chi$:
\begin{eqnarray}
	\nonumber
	&&H[\psi,\chi; \tau,\{ \delta \tau\}] = \int d^{2}r \, \Big[
	-\frac{1}{2}\overline{\psi}({\bf r})\hat{\partial} \psi({\bf r}) \; -\frac{1}{2}\overline{\chi}({\bf r})\hat{\partial} \chi({\bf r}) \\
	\nonumber
	\\
	&-& \frac{1}{2}\bigl(\tau+\delta \tau_1({\bf r})\bigr) \, \overline{\psi}({\bf r})\psi({\bf r})
	- \; \frac{1}{2}\bigl(\tau+\delta \tau_2({\bf r})\bigr)\, \overline{\chi}({\bf r})\chi({\bf r}) \, \nonumber\\&+& \,
	g_0(\overline{\psi}({\bf r})\psi({\bf r}))(\overline{\chi}({\bf r})\chi({\bf r}))
	\Big] \,,
	\label{6}
\end{eqnarray}
where $\delta \tau_1({\bf r})$ and $\delta \tau_2({\bf r})$  are spatially uncorrelated  Gaussian variables with zero mean, i.e., $\overline{ \delta \tau_1 ({\bf r}) } = 0$ and $\overline{ \delta \tau_2 ({\bf r})} = 0$, and the covariance function
\begin{equation}
	\label{7}
	\overline{ \delta \tau_i({\bf r})\delta \tau_j({\bf r}') }  \; = \; 4 u_{0} \, \delta_{ij}\delta({\bf r} - {\bf r}') \,,
\end{equation}
with $\delta_{ij}$ being the Kronecker-symbol, such that $\delta_{ij} = 1$ for $i = j$, and zero, otherwise, and
the parameter $u_{0} \ll 1$ defines, as above, the strength of disorder.

For a given realization of the random terms $\delta \tau_1 ({\bf r})$ and  $\delta \tau_2 ({\bf r}) $, the partition function of the continuous-space system under study is
\begin{equation}
	\label{8}
	Z[\tau; \{\delta \tau\}] \; = \; \int {\cal D} \psi \; \exp\bigl\{-H[\psi; \tau, \{\delta \tau\} ]\bigr\} \; = \;
	\exp\bigl\{ - F[\tau; \{\delta \tau\}]\bigr\} \,,
\end{equation}
where $F[\tau; \{\delta \tau\}]$ is a random free-energy function. The internal energy of the considered system is given by the first derivative of the free energy over the parameter (temperature) $``\tau"$:
\begin{equation}
	\label{9}
	E[\tau; \{\delta \tau\}] \; = \; \frac{\partial}{\partial\tau} F[\tau;\{\delta \tau\}]] \,.
\end{equation}
We note that, first, $E[\tau;\{\delta \tau\}]$ must be a singular function of $\tau$ in the limit $\tau \to 0$. Second, it is a {\it random function} that  exhibits sample-to-sample fluctuations. The probability density function of these fluctuations is the focal point of the present study. Thirdly, our conclusions on a lack of self-averaging of $E[\tau;\{\delta \tau\}]$ will imply directly, in view of Eq. (\ref{9}) that $F[\tau;\{\delta \tau\}]] $ is not self-averaging too.

Four decades ago the model~\eqref{6} was studied in Ref.~\cite{DD84}. In terms of the above spinor representation, it was shown that for $g_{0} > 0$ (when the specific heat of the pure model is divergent), in the vicinity of the critical point the specific heat of the random system exhibits a double-logarithmic singularity $C_{rand}(\tau) \, \sim \, \ln\ln\bigl(1/|\tau|\bigr)$, i.e., it exhibits precisely the same behavior as one observes for a disordered Ising model. Conversely, for $g_{0}<0$, (when the specific heat of the pure system remains finite), the specific heat of the random system also remains finite but shows a cusp singularity of the form $C_{rand}(\tau) \sim \bigl[\ln\ln\bigl(1/|\tau|\bigr)\bigr]^{-1}$. This prediction indicates the breakdown of Harris criterion in the disordered Baxter model. We will discuss this issue later.

\section{Replica formalism}
\label{replica}

In this section, we implement the replica formalism to perform the averaging over disorder. For this purpose, we rewrite formally Eq.~\eqref{9} as
\begin{equation}
	\label{10}
	E[\tau; \delta\tau] \; = \; \lim_{\epsilon \to 0} \;
	\frac{1}{\epsilon}\bigl(F[\tau + \epsilon; \delta\tau] \; - \; F[\tau; \delta\tau]\bigr) \,.
\end{equation}
Supposing next that $\epsilon$ is finite (it has to be set to zero in the final result), we rearrange the latter expression to get
\begin{equation}
	\label{11}
	\epsilon \, E[\tau; \delta\tau] \; = \;
	F[\tau + \epsilon; \delta\tau] \; - \; F[\tau; \delta\tau] \,.
\end{equation}
Further on, in virtue of the definition of the free energy, Eq.(\ref{8}), the above relation can be expressed in terms of the ratio of two partition functions at two different temperatures. Namely, we have
\begin{equation}
	\label{12}
	\exp\bigl\{-\epsilon \, E[\tau; \delta\tau]\bigr\} \; = \;
	Z[\tau+\epsilon; \delta\tau] \, Z^{-1}[\tau; \delta\tau] \,.
\end{equation}
Taking the $N$-th power of both sides of the above equation, and then performing averaging over disorder, we obtain
\begin{equation}
	\label{13}
	\int \, dE \; P_{\tau} (E) \; \exp\bigl\{-\epsilon N \, E\bigr\} \; = \;
	\overline{ Z^{N}[\tau+\epsilon; \delta\tau] \, Z^{-N}[\tau; \delta\tau] } \,,
\end{equation}
where $P_{\tau} (E)$ is the probability distribution function of the internal energy of the system
at a given value of the parameter $\tau$. Using next the standard tricks of the replica
formalism, we express the above relation as
\begin{eqnarray}
	\label{14}
	\int \, dE \; P_{\tau} (E) \; \exp\bigl\{-\epsilon N \, E\bigr\} \; &=& \; \lim_{M\to 0}
	\overline{ Z^{N}[\tau+\epsilon; \delta\tau] \, Z^{M-N}[\tau; \delta\tau]} \; \nonumber\\ &\equiv& \;
	\lim_{M\to 0} \; {\cal Z}(M, N; \tau, \epsilon) \,.
\end{eqnarray}
In doing so,  it is first assumed that both $M$ and $N$ are integers, such that $M > N$. Then,  after determining ${\cal Z}(M, N; \tau, \epsilon)$ for arbitrary integer $M$ and $N$, these parameters are analytically continued to arbitrary real values, and then eventually the limit $M\to 0$ is taken. After that, introducing a parameter $s = \epsilon N$ and taking the limit $\epsilon \to 0$ (provided that such a limit exists) the relation~\eqref{14} takes the form of the Laplace transform of the probability distribution function $P_{\tau} (E)$:
\begin{eqnarray}
	\label{15}
	\int \, dE \; P_{\tau} (E) \; \exp\bigl\{-s \, E\bigr\} \; &=& \;
	\lim_{\epsilon\to 0} \lim_{M\to 0} \;  {\cal Z}(M, s/\epsilon; \tau, \epsilon) \; \nonumber\\ &\equiv& \;
	\tilde{{\cal Z}} (s, \tau)\; \equiv \;\exp\{-{\cal F}(s,\tau)\} \,.
\end{eqnarray}
We note parenthetically that, in view of the form of the latter expression, the free energy ${\cal F}(s,\tau)$ taken with the opposite sign of $s$ can be interpreted as the cumulant-generating function of the probability density function $P_{\tau} (E)$. Therefore, all the cumulants of $E$ can be obtained by a \textit{mere} differentiation of ${\cal F}(s,\tau)$ and then setting $s = 0$. Accordingly, the moments of $E$ are found by differentiating  $\exp\{-{\cal F}(s,\tau)\}$ with respect to $s$, and then setting $s = 0$.

As a first step, we consider the structure of the replica partition function ${\cal Z}(M, N; \tau, \epsilon)$. According to the definitions~\eqref{8} and \eqref{14},
\begin{equation}
	\label{17}
	{\cal Z}(M, N; \tau, \epsilon) \; = \;
	\int {\cal D}\psi \; \overline{ \left( \exp\Bigl\{
		-\sum_{a=1}^{N} H[\psi_{a}; \tau+\epsilon, \delta\tau] \; -
		\sum_{a=N+1}^{M} H[\psi_{a}; \tau, \delta\tau]
		\Bigr\} \right)} \,.
\end{equation}
Inserting in the latter expression the Hamiltonian~\eqref{6} and performing simple Gaussian averaging
over $\delta\tau({\bf r})$ (using Eq.~\eqref{7}), we get:
\begin{equation}
	\label{18}
	{\cal Z}(M, N; \tau, \epsilon) \; = \;  \int {\cal D}\psi \; \exp\Bigl\{- {\cal H}_{M,N}[\psi; \tau, \epsilon]\Bigr\}
	\; \equiv \; \exp\bigl\{- {\cal F}(M,N;\tau,\epsilon)\bigr\} \,,
\end{equation}
where ${\cal F}(M,N;\tau,\epsilon)$ is called the ``replica free energy" and ${\cal H}_{M,N}[\psi; \tau, \epsilon]$ is the \textit{effective} Hamiltonian with replica-dependent masses:
\begin{eqnarray}
	\nonumber
	{\cal H}_{M,N}[\psi, \chi; \tau, \epsilon] 
	&=&
	\int d^{2}r \, 
	\Biggl[
	- \frac{1}{2}\sum_{a=1}^{M}
	\bigl(\overline{\psi}_{a}({\bf r})\hat{\partial} \psi_{a}({\bf r}) 
	+ \overline{\chi}_{a}({\bf r})\hat{\partial} \chi_{a}({\bf r})\bigr)
	\; \nonumber\\&-& \;
	\frac{1}{2}\sum_{a=1}^{M} 
	m_{a} \bigl(\overline{\psi}_{a}({\bf r})\psi_{a}({\bf r})
	+ \overline{\chi}_{a}({\bf r})\chi_{a}({\bf r})\bigr) 
	\nonumber
	\\
	&-& 
	\frac{1}{2} u_{0} \sum_{a,b=1}^{M} \Bigl(\bigl(\overline{\psi}_{a}({\bf r})\psi_{a}({\bf r})\bigr)
	\bigl(\overline{\psi}_{b}({\bf r})\psi_{b}({\bf r})\bigr) \nonumber\\&+& \bigl(\overline{\chi}_{a}({\bf r})\chi_{a}({\bf r})\bigr)
	\bigl(\overline{\chi}_{b}({\bf r})\chi_{b}({\bf r})\bigr)\Bigr)
	\nonumber\\
	&+&
	g_{0} \sum_{a=1}^{M} \bigl(\overline{\psi}_{a}({\bf r})\psi_{a}({\bf r})\bigr)
	\bigl(\overline{\chi}_{a}({\bf r})\chi_{a}({\bf r})\bigr)
	\, \nonumber\\ &+& \,                                    
	\gamma \sum_{a,b=1}^{M} \bigl(\overline{\psi}_{a}({\bf r})\psi_{a}({\bf r})\bigr)
	\bigl(\overline{\chi}_{b}({\bf r})\chi_{b}({\bf r})\bigr)
	\Biggr] \,,
	\label{19}
\end{eqnarray}
where
\begin{equation}
	\label{20}
	m_{a} \; = \; \left\{
	\begin{array}{ll}
		(\tau+\epsilon)\, \; \; \mbox{for} \; a = 1, ..., N \, ,
		\\
		\\
		\tau \, \; \; \; \; \; \;  \; \; \; \; \, \mbox{for} \; a = N+1,
		..., M \, .
	\end{array}
	\right.
\end{equation}
The last term in Eq.~\eqref{19} does not directly appear after averaging over quenched disorder, nevertheless we include it, since it occurs in the course of renormalization. Later on, we will derive the function ${\cal F}(M, N; \tau, \epsilon)$, Eq.~\eqref{18}, using the standard procedure of the RG technique.

\section{Renormalization group flow equations}
\label{RG}

It is well known that in two dimensions the spinor field theory with the four-fermion interactions is renormalizable (see e.g., ~\cite{DD_rev83}). The renormalization of the replica Hamiltonian~\eqref{19} can be done in a standard way by integrating out the short wavelength degrees of freedom in the band $\tilde{\Lambda} < p < \Lambda$, where $\Lambda$ and $\tilde{\Lambda}$ are the old and the new ultraviolet momenta $p$ cut-offs, respectively. One can then show according to the Wilson's RG approach that the renormalization of the charges $g$, $u$, $\gamma$ and of the mass $m_{a}$ in the replicated Hamiltonian~\eqref{19} is given by the following set of flow equations (see, e.g.,~\cite{DD84}):
\begin{eqnarray}
	\label{21}
	\frac{d}{d\xi} g(\xi) &=& -\frac{4}{\pi}\; g(\xi) u(\xi) \,,
	\\
	\nonumber
	\\
	\label{22}
	\frac{d}{d\xi} u(\xi) &=& -\frac{2}{\pi}(2 - M) \; u^{2}(\xi)+\frac{2}{\pi} M \; \gamma^{2}(\xi)+\frac{4}{\pi} \; g(\xi) \gamma(\xi) \,,
	\\
	\nonumber
	\\
	\label{23}
	\frac{d}{d\xi} \gamma(\xi) &=& \frac{4}{\pi} \; g(\xi) u(\xi)-\frac{4}{\pi} (1-M)\; u(\xi) \gamma(\xi) \,,
	\\
	\nonumber
	\\
	\frac{d}{d\xi} m_{a}(\xi) &=& -\frac{2}{\pi}\Bigl(m_{a}(\xi) \; - \; \sum_{b=1}^{M} m_{b}(\xi)\Bigr) \; u(\xi)\nonumber\\&+&\frac{2}{\pi} \sum_{b=1}^{M} m_{b}(\xi) \gamma(\xi) +\frac{2}{\pi}m_{a}(\xi) g(\xi) \,,
	\label{24}
\end{eqnarray}
where $\xi = \ln\bigl(\Lambda/\tilde{\Lambda}\bigr)$ and
\begin{equation}
	\label{25}
	m_{a}(\xi) \; = \; \left\{
	\begin{array}{ll}
		\tilde{m}(\xi)\,
		\; \;
		\mbox{for} \; a = 1, ..., N \; ,
		\\
		\\
		m(\xi) \,
		\; \;
		\mbox{for} \; a = N+1, ..., M \; ,
	\end{array}
	\right.
\end{equation}
with the initial conditions at the lattice scale: $g(\xi\sim 1) =  g_{0}$,  $u(\xi\sim 1) =  u_{0}$,  $\gamma(\xi\sim 1) =  0$, $\; \tilde{m}(\xi\sim 1) = (\tau+\epsilon)$ and $m(\xi\sim 1) = \tau$. Denoting $x= \frac{4}{\pi}\xi$ and substituting Eq.~\eqref{25} into Eq.~\eqref{24}, we find that in the limit $M\to 0$ the above equations are simplified as,
\begin{eqnarray} 
	\label{26}
	\frac{d}{d x} g(x) &=& - g(x) u(x) \,,
	\\
	\nonumber
	\\
	\label{27}
	\frac{d}{d x} u(x) &=& -u^{2}(x)+ \; g(x) \gamma(x) \,,
	\\
	\nonumber
	\\
	\label{28}
	\frac{d}{d x} \gamma(x) &=& \; g(x) u(x)-\; u(x) \gamma(x) \,,
	\\
	\nonumber
	\\ 
	\label{29}
	2\frac{d}{d x} \tilde m (x) &=& -\Bigl((1-N)\tilde m(x) + N m(x)\Bigr) \; u(x)\nonumber\\&+& N(\tilde m(x) -m(x))\gamma(x) + \tilde m (x) g(x) \,, 
	\\
	\nonumber
	\\ 
	\label{30}
	2\frac{d}{d x}  m (x) &=& -\Bigl((1+N) m(x) - N \tilde m(x)\Bigr) \; u(x)\nonumber\\&+&N (\tilde m(x) -m(x))\gamma(x) + m (x) g(x)  \,.
\end{eqnarray}

Setting $\tilde m(x)=m(x)+\epsilon \Delta (x)$ (so that $\Delta(x\sim 1) = 1$) and $\epsilon N = s$, we find that Eqs.~\eqref{29} and \eqref{30} reduce to
\begin{eqnarray}
	\label{34}
	2\frac{d}{d x} \Delta (x) &=& -\left( u(x)- g(x)\right)  \Delta (x)  \,,
	\\
	\nonumber
	\\ \label{35}
	2\frac{d}{d x}  m (x) &=& -\left( u(x)- g(x)\right)  m (x) + s (u(x)+\gamma(x)) \Delta(x) \; .
\end{eqnarray}

Finally, the set of RG equations~\eqref{26},\eqref{27}, \eqref{28}, \eqref{34}, and \eqref{35} are solved with the following initial conditions:
\begin{eqnarray} 
	\label{261}
	g(x=1) &=& g_0 \,, u(x=1) = u_0 \,, \gamma(x=1) = 0 \,,
	\\
	\nonumber
	\\
	\label{271}
	\Delta(x=1) &=& 1 \,, m(x=1) = \tau \,.
\end{eqnarray}
Notice that Eqs.~\eqref{26}-\eqref{28} are independent of Eqs.~\eqref{34}-\eqref{35}, and the solutions of Eqs.~\eqref{34}-\eqref{35} are defined by the solutions of Eqs.~\eqref{26}-\eqref{28}.

The solutions of Eqs.~\eqref{27} and \eqref{28} in terms of $g$ are:
\begin{eqnarray} 
	\label{262}
	u(g) = \vert g \vert \sqrt{\ln^2 \left(\frac{g}{g_0} \right) + \left( \frac{u_0}{g_0} \right)^2} \,,
	\\
	\nonumber
	\\
	\label{272}
	\gamma(g) = g \ln \left(\frac{g_0}{g} \right) \,.
\end{eqnarray}
Substituting the solution of $u$~\eqref{262} in Eq.~\eqref{26}, we get
\begin{equation} 
	\label{263}
	\frac{d}{d x} g(x) = - g ~\vert g \vert \sqrt{\ln^2 \left(\frac{g}{g_0} \right) + \left( \frac{u_0}{g_0} \right)^2} \,.
\end{equation}
We observe a pronounced dependence on the sign of the coupling parameter $g_0$, indicating that the regimes $g_0 < 0$ and $g_0 > 0$ must be treated separately in Eq.~\eqref{263}. In particular, the case $g_0 < 0$, which corresponds to a frustrated interaction, is substantially more intricate~\footnote{For $g_0 < 0$, the factor $g \vert g \vert$ in Eq.~\eqref{263} changes the structure of the RG trajectory. Consequently, the parametrization used for $g_0 > 0$ does not provide a straightforward continuation to this frustrated regime. The sign change would also modify the combinations $u-g$ and $u+\gamma$ entering the flow equations for $\Delta(x)$ and $m(x)$, which are the quantities needed to compute the moments of the critical energy. This makes the derivation of compact closed expressions substantially more intricate. We therefore restrict the analytical treatment to $g_0>0$ and address the case $g_0<0$ numerically.} than the ferromagnetic-like regime $g_0 > 0$ and will be analyzed in detail elsewhere.

In the present work, we therefore restrict the analytical treatment to the case $g_0 > 0$, while the frustrated regime will be addressed only through numerical simulations. For positive $g_0$ Eqs.~\eqref{26}–\eqref{28} yield the following solutions:
\begin{eqnarray} 
	\label{264}
	\gamma(x) &=& \frac{g_0}{\phi(x)} \ln \left( \phi(x) \right) \,,
	\\
	\nonumber
	\\
	\label{274}
	u(x) &=& \frac{g_0}{\phi(x)} \sqrt{\ln^2 \left[ \phi(x) \right] + \lambda^2} \,,
	\\
	\nonumber
	\\
	\label{284}
	g(x) &=& \frac{g_0}{\phi(x)} \,,
\end{eqnarray}
where the parameter $\lambda = \frac{u_0}{g_0}$ and the function $\phi(x)$ is defined by the following equation:
\begin{equation}
	\label{285}
	\int_{1}^{\phi(x)} \; \frac{ d\xi}{\sqrt{\ln^2 \xi + \lambda^2}} \; = \; g_0 (x-1) \; ,
\end{equation}
which further simplifies as
\begin{equation}
	\label{286}
	\frac{d \phi(x)}{d x} \; = \; g_0 \sqrt{\ln^2 (\phi) + \lambda^2} \; ,
\end{equation}
with initial condition:
\begin{equation}
	\label{287}
	\phi(x = 1) = 1 \; .
\end{equation}
With the above solutions~\eqref{264}-\eqref{284} in hand, the solutions of Eqs.~\eqref{34} and \eqref{35} are given as
\begin{eqnarray} 
	\label{288}
	\Delta(\phi) \; &=& \; \frac{1}{\sqrt{\phi}} \exp \left\{ \frac{1}{2} \, {\rm arcsinh} \left( \frac{1}{\lambda} \ln \phi  \right)  \right\} \,,
	\\
	\nonumber
	\\
	\label{289}
	m(\phi, s) \; &=& \; \frac{1}{\sqrt{\phi}} \exp \left\{ \frac{1}{2} \, {\rm arcsinh} \left( \frac{1}{\lambda} \ln \phi  \right)  \right\} \;\nonumber\\
	&\times& \left[  \frac{1}{2} \; s \ln(\phi) \; + \; \frac{1}{2} \; s \sqrt{\ln^2 \phi \; + \; \lambda^2} \; - \; \frac{1}{2} \lambda \; s \; + \; \tau  \right] \; .
\end{eqnarray}

\section{Critical internal energy}
\label{moments}

According to the arguments presented in Sec.~\ref{replica}, for a given temperature parameter $\tau$ the internal energy probability distribution function $P_{\tau}(E)$ is defined by the relation~\eqref{15},
\begin{equation}
	\label{45}
	\int \, dE \; P_{\tau} (E) \; \exp\bigl\{-s \, E\bigr\} \; = \; 
	\exp\{-{\cal F}(s,\tau)\}
	\; ,
\end{equation}
where
\begin{equation}
	\label{46}
	{\cal F}(s,\tau) \; = \; -\lim_{\epsilon\to 0} \lim_{M\to 0} \,
	\ln\bigl[ {\cal Z}(M, N; \tau, \epsilon)\bigr]\Big|_{N=s/\epsilon}
	\; ,
\end{equation}
and the replica partition function ${\cal Z}(M, N; \tau, \epsilon)$  is given by Eqs.~\eqref{18}-\eqref{19}. In the standard way, the singular (critical) part of the corresponding replica free energy
\begin{equation}
	\label{47}
	{\cal F}(0,N;\tau,\epsilon) \; = \; - \ln\bigl[{\cal Z}(0,N;\tau,\epsilon)\bigr]
\end{equation}
can be computed by the renormalization group methods~\cite{Larkin-Khmelnitskii,Aharony} (see also \cite{DD_rev83}):
\begin{equation}
\hspace*{-1.9cm}
	\label{48}
	{\cal F}(0,N;\tau,\epsilon) \, = \,  - L^{2} \, \lim_{M\to 0} \, 
	\int_{0}^{p_{0}} \frac{d^{2}p}{(2\pi)^{2}} \;
	\ln\Biggl[
	\det\Bigl(i\hat{p} + \tilde{m}(p) \, \hat{1}\Bigr)^{N} \times
	\det\Bigl(i\hat{p} + m(p) \, \hat{1}\Bigr)^{M-N}
	\Biggr]
\end{equation}
where $L$ is the size of the system, $\hat{p} = \hat{\sigma}_{1} p_{x} + \hat{\sigma}_{2} p_{y}$ (cf.~Eqs.~\eqref{2}--\eqref{3}) and $\hat{1}$ is the unit matrix. Here $m(p)$ and $\tilde{m}(p) = m(p) + \Delta(p)$ are renormalized mass parameters which are considered to be dependent on the scale according to Eqs.~\eqref{34},\eqref{35} and \eqref{26}-\eqref{28} with $x = \frac{4}{\pi} \xi =  \frac{4}{\pi}\ln(1/p)$. As the renormalization parameter $x$ is considered to be bounded by the condition $x \gtrsim 1/g_{0}$ the upper limit in the integration over momenta $p$ in Eq.~\eqref{48} is
\begin{equation}
	\label{48a}
	p_{0} \sim \exp\{-1/g_{0}\} \; .
\end{equation}
Simple calculations yield
\begin{eqnarray}
	\nonumber
	{\cal F}(0,N;\tau,\epsilon) &=& - L^{2} \int_{0}^{p_{0}} \frac{d^{2}p}{(2\pi)^{2}} \;
	\Biggl[
	N \ln\bigl(p^{2} + \tilde{m}^{2}(p)\bigr) -  N \ln\bigl(p^{2} + m^{2}(p)\bigr)
	\Biggr]
	\\
	\nonumber
	\\
	\nonumber
	&=& -\frac{1}{2\pi} L^{2} N \int_{0}^{p_{0}} dp \, p \;
	\ln\Biggl[
	\frac{p^{2} + \bigl(m(p) + \epsilon \Delta(p)\bigr)^{2}}{p^{2} + m^{2}(p)}
	\Biggr]
	\\
	\nonumber
	\\
	\nonumber
	&\simeq& -\frac{1}{2\pi} L^{2} N \int_{0}^{p_{0}} dp \, p \;
	\ln\Biggl[
	1 \; + \; \epsilon \, \frac{2 m(p) \Delta(p)}{p^{2} + m^{2}(p)}
	\Biggr]
	\\
	\nonumber
	\\ 
	&\simeq& -\frac{1}{\pi} L^{2} (\epsilon N) \int_{0}^{p_{0}} dp \, p \;
	\frac{ m(p) \Delta(p)}{p^{2} + m^{2}(p)} \; + \; \mbox{O}\bigl(\epsilon^{2}\bigr)
	\label{49} 
\end{eqnarray}
Substituting here $(\epsilon N) \, = \, s$, according to Eq.~\eqref{46}, and then performing a change of variable $p \rightarrow x$, we get
\begin{equation}
	\label{501}
	{\cal F}(s,\tau) \; = \; -\frac{1}{\pi} L^{2} s \int_{1}^{\infty} dx \; \frac{ m(x) \Delta(x)}{1 + {\rm e}^{2x} \; m^{2}(x)} \; .
\end{equation}
where $\Delta$ and $m$ are given by Eqs.~\eqref{288}-\eqref{289}. Note that although our solutions in Eqs.~\eqref{288} and~\eqref{289} are formally valid only for $x > 1/g_0$ we extend these solutions down to $x=1$, at which point the initial conditions are specified (see Eq.~\eqref{271}). This is, of course, an uncontrollable approximation which \textit{tacitly} assumes that the solutions are slowly varying functions of the integration variable. Therefore,  given that the kernels are positive definite, Eq.~\eqref{501} provides an upper bound on the actual behavior of the free energy ${\cal F}$.

As discussed earlier, all the moments of the internal energy $E$ can be found by differentiating the function $\exp\{-{\cal F}(s,\tau)\}$ with respect to $s$, and then substituting $s = 0$. We calculate the first two moments below as:
\begin{eqnarray}
	\nonumber
	\overline{  E } \;
	&=& \; - \frac{\partial}{\partial s} \; \exp\{-{\cal F}(s,\tau)\} \; \Bigg\vert_{s=0} \; ,
	\\
	\nonumber
	\\
	\label{502}
	&=& -\frac{L^{2}}{\pi} \int_{1}^{\infty} dx \; \frac{ m(x) \Delta(x)}{1 + {\rm e}^{2x} \; m^{2}(x)}  \; \Bigg\vert_{s=0} \; ,
\end{eqnarray}
and
\begin{eqnarray}
	\nonumber
	\overline{  E^2 } \; &=& \; - \frac{\partial^2}{\partial s^2} \; \exp\{-{\cal F}(s,\tau)\} \; \Bigg\vert_{s=0} \; ,
	\\
	\nonumber
	\\
	\label{503}
	&=& \;  \overline{E}^2 - \frac{\partial^2}{\partial s^2} \; {\cal F}(s,\tau) \; \Bigg\vert_{s=0} \; .
\end{eqnarray}
The above expression gives the variance of the internal energy probability distribution as
\begin{equation}
	\label{504}
	\overline{E^2} - \overline{E}^2 \; = \; \frac{\partial^2}{\partial s^2} \left[ \frac{L^{2}}{\pi} s \int_{1}^{\infty} dx \; \frac{ m(x) \Delta(x)}{1 + {\rm e}^{2x} \; m^{2}(x)} \right] \; \Bigg\vert_{s=0} \; .
\end{equation}
Note that the quantity $m(x)$ implicitly depends on parameter $s$, $m(x) \equiv m(x,s)$; see Eq.~\eqref{289}. For simplification, changing variable $x \rightarrow \phi$ using Eq.~\eqref{286} as
\begin{equation}
	\label{505}
	x(\phi) \; = \; \frac{1}{g_0} \int_{1}^{\phi} \; \frac{ d\xi}{\sqrt{\ln^2 \xi + \lambda^2}} \; + \; 1 \; ,
\end{equation}
we obtain
\begin{equation}
	\label{506}
	\overline{E} \; = \; -\frac{L^{2} \tau}{\pi g_0} \int_{1}^{\infty} d\phi \; \frac{\Delta^2 (\phi)}{\sqrt{\ln^2 \phi + \lambda^2} \, \left( \, 1 + \tau^2 \Delta^2(\phi) \, {\rm e}^{2 \, x(\phi)} \, \right)}  \; ,
\end{equation}
and
\begin{equation}
\hspace*{-1.2cm}
	\label{507}
	\overline{E^2} - \overline{E}^2 \; = \; \frac{L^{2}}{\pi g_0} \int_{1}^{\infty} d\phi \; \frac{  \left( \, 1 - \tau^2 \Delta^2(\phi) \, {\rm e}^{2 \, x(\phi)} \, \right) \, \Delta^2 (\phi) \, \left( \ln \phi + \sqrt{\ln^2 \phi + \lambda^2} - \lambda \right)}{\sqrt{\ln^2 \phi + \lambda^2} \; \left( \, 1 + \tau^2 \Delta^2(\phi) \, {\rm e}^{2 \, x(\phi)} \, \right)^2}  \; .
\end{equation}
Clearly, the relative variance~\eqref{relVar} of the critical internal energy can be obtained by using Eqs.~\eqref{506} and \eqref{507}, with solution of $\Delta$ from Eq.~\eqref{288} and relation~\eqref{505}.

\begin{figure}[t!]
\centering
	\includegraphics[width=0.75\textwidth]{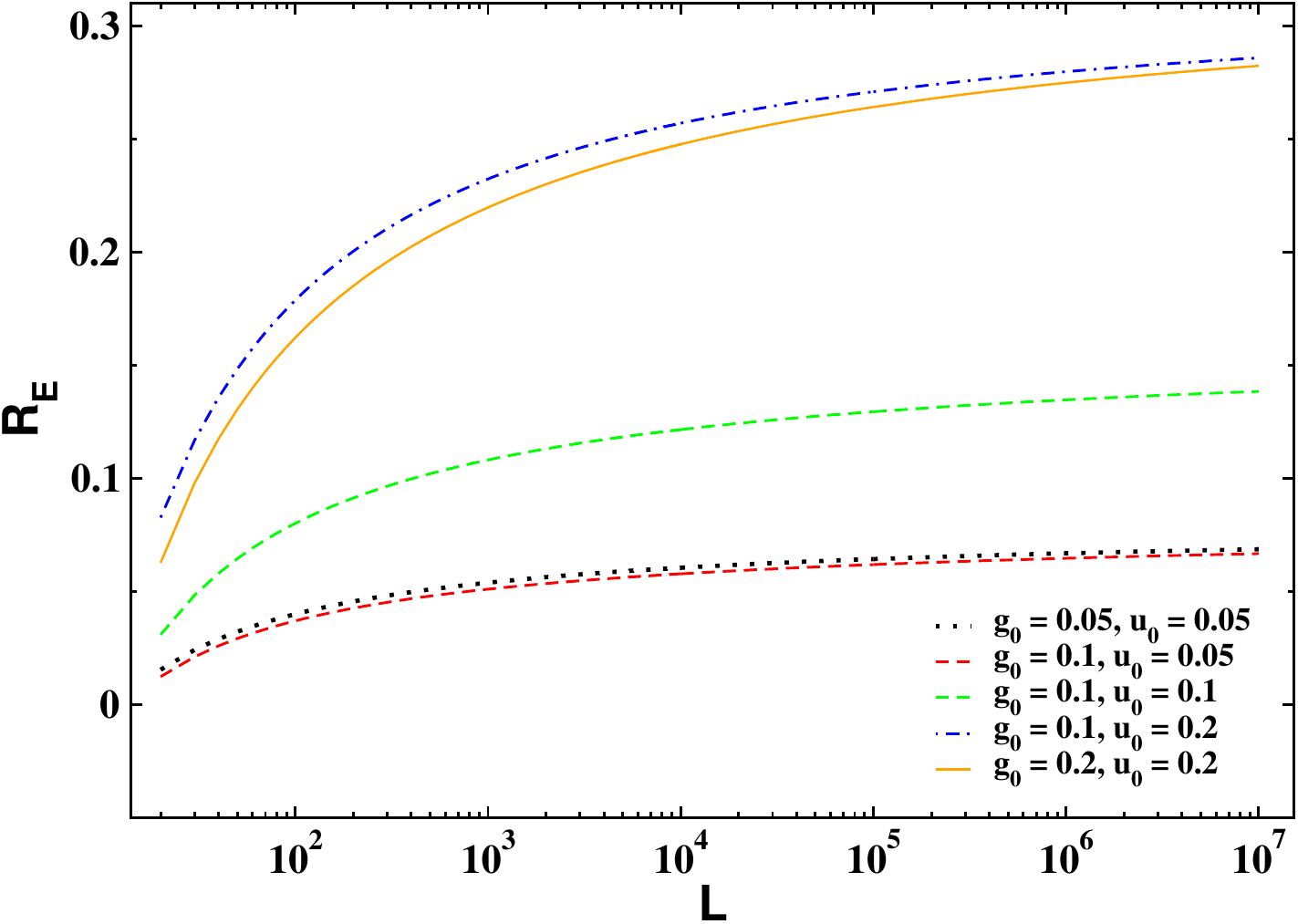}
	\caption{Plot of the relative variance $R_E$ as function of linear system size $L$ on a log-linear scale for different values of parameters $g_0$ and $u_0$ (see the keys).}
	\label{Fig_Re}
\end{figure}

In the above expressions, the reduced temperature parameter $\tau$ which locates the {\it pseudo-critical point} $\tau_{*}$ is chosen as follows. For large but finite system size $L$, $\tau_{*}$ is defined by the condition that the correlation length $\xi_{\rm c}(\tau)$ becomes of the order of $L$. In the {\it pure} Baxter model the correlation length $\xi_{\rm c}(\tau) \sim \tau^{-\nu}$ is defined by non-universal critical exponent $\nu = 1 - \frac{2}{\pi} g_{0}$ being dependent on the coupling parameter $g_{0}$. However, it is argued~\cite{DD84} that, in the {\it disordered} Baxter model, the RG trajectories flow to the universal Ising-model fixed point. In this case, $\nu = 1$ (see also the discussion in Sec.~\ref{numerical}), and consequently the pseudo-critical point in the disordered Baxter model can be estimated as
\begin{equation}
	\label{51}
	\tau_{*}(L) \, \sim \, L^{-1}
	\; .
\end{equation}

In Fig.~\ref{Fig_Re}, we plot the relative variance $R_E$  defined in Eq.~\eqref{relVar}, obtained by numerically integrating Eqs.~\eqref{506}–\eqref{507}. We fix the critical exponent to $\nu=1$. For different positive values of $g_0$ and $u_0$, the quantity $R_E$ increases with the system linear size $L$ and eventually approaches a finite constant at large $L$. This behavior clearly demonstrates the absence of self-averaging of the critical internal energy. We emphasis that while different curves in Fig.~\ref{Fig_Re} exhibit a noticeable dependence on $u_0$ and a relatively weaker dependency on $g_0$, the central result is the robust non-vanishing asymptotic value of $R_E$.

Now let us also briefly discuss the internal energy probability distribution function $P_{\tau} (E)$~\eqref{45} at $\tau_{*}$. Formally, we can reconstruct the function $P_{\tau} (E)$ by the inverse Laplace transform:
\begin{equation}
	\label{16}
	P_{\tau} (E) \; = \; \int_{-i\infty}^{+i\infty}\frac{ds}{2\pi i} \; \exp\{-{\cal F}(s,\tau)\} \exp\bigl\{ s \, E \bigr\} \; ,
\end{equation}
where function
\begin{equation}
	\label{52}
	\exp\{-{\cal F}(s,\tau)\} \, = \, \exp\biggl\{
	\frac{1}{\pi g_0} L^{2} s \int_{1}^{\infty} d\phi \, \frac{ m(\phi) \Delta(\phi)}{\sqrt{\ln^2 \phi + \lambda^2} \, \left(1 + {\rm e}^{2 \, x(\phi)} \, m^{2}(\phi) \right)} \biggr\} \;
\end{equation}
is obtained with the help of Eq.~\eqref{501} and a change of variable $x \rightarrow \phi$ using Eq.~\eqref{505}. Again, the solutions $\Delta(\phi)$ and $m(\phi)$ are defined by Eqs.~\eqref{288}-\eqref{289}.

\section{Numerical results}
\label{numerical}

In this section, we numerically investigate the moments of the critical internal energy (see Appendix~\ref{2DRIM} for the disordered Ising model) for both positive and negative $g_0$. For this purpose, we run extensive simulations of the disordered Baxter model~\eqref{ham} on finite size lattices.

We consider the model~\eqref{ham} with quenched bond variables $\{ J_{ij}^{(1)}, \, J_{kl}^{(2)} \}$ of the same form as in Sec.~\ref{II}: $J_{ij}^{(1)} = J + \delta J_{ij}^{(1)}$ and $J_{kl}^{(2)} = J + \delta J_{kl}^{(2)}$. In the simulations, we fix the constant $J$ to unity, while the fluctuations $\delta J_{ij}^{(1)}$ and $\delta J_{kl}^{(2)}$ are drawn from a Gaussian distribution~\eqref{gauss} with zero mean and variance $u_0$ (consistent with our analytical treatment). Due to the Gaussian nature of the fluctuations, the couplings $\{ J_{ij}^{(1)}, \, J_{kl}^{(2)} \}$ can also take negative values, leading to frustration. To equilibrate such a frustrated system at the critical point, we employ an optimized parallel tempering (PT) technique~\cite{Hukushima96,Marinari92}. For statistical accuracy, we consider approximately $1000$--$2000$ disorder samples for each system size $L \in [12, 96]$. For each disorder sample, $100$ independent configurations are generated for thermal averaging. The statistical error bars on different observables are estimated using Jackknife resampling technique~\cite{efron1982jackknife}.

\begin{figure}[t!]
\centering
	\includegraphics[width=0.48\textwidth]{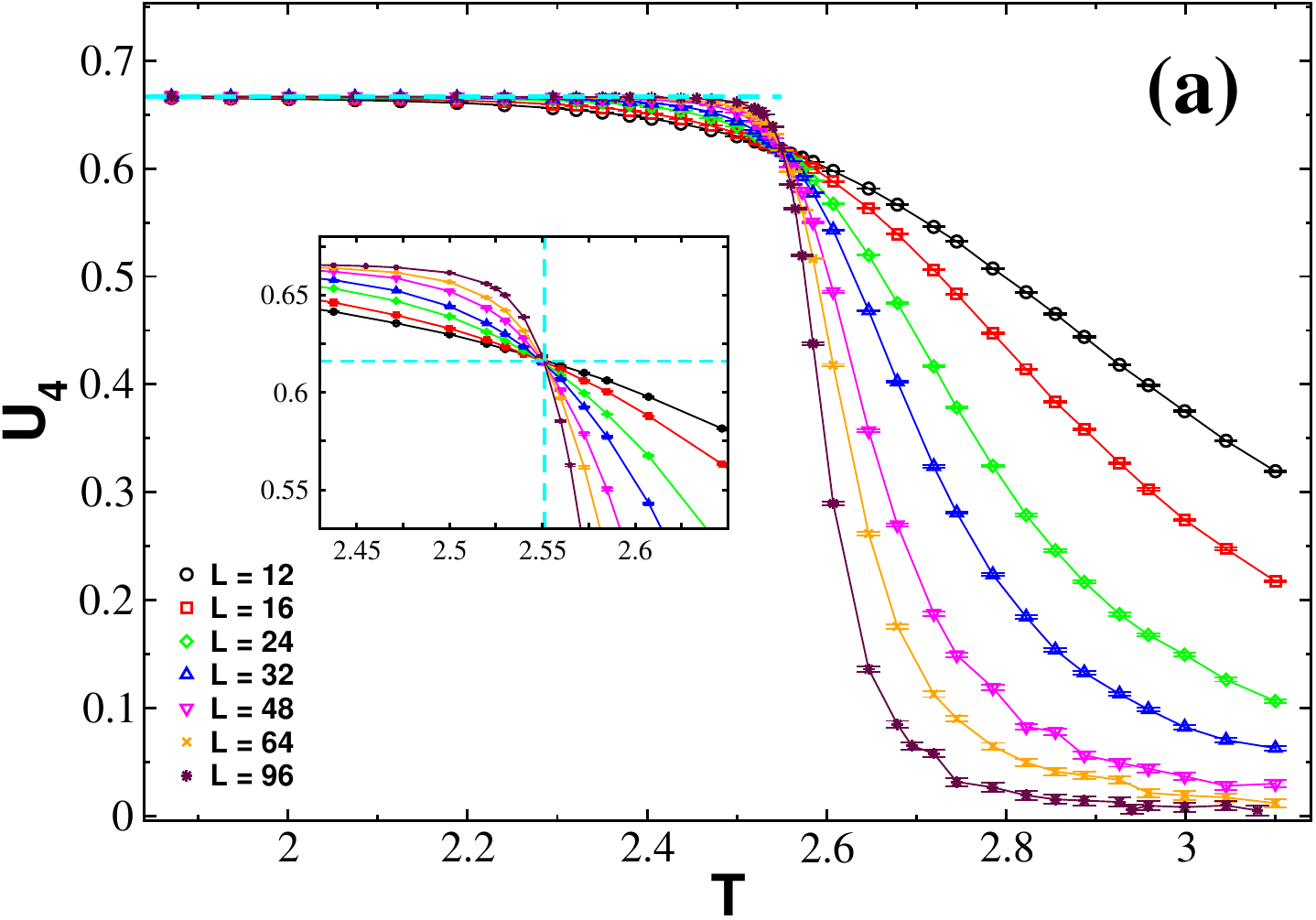}
	\includegraphics[width=0.48\textwidth]{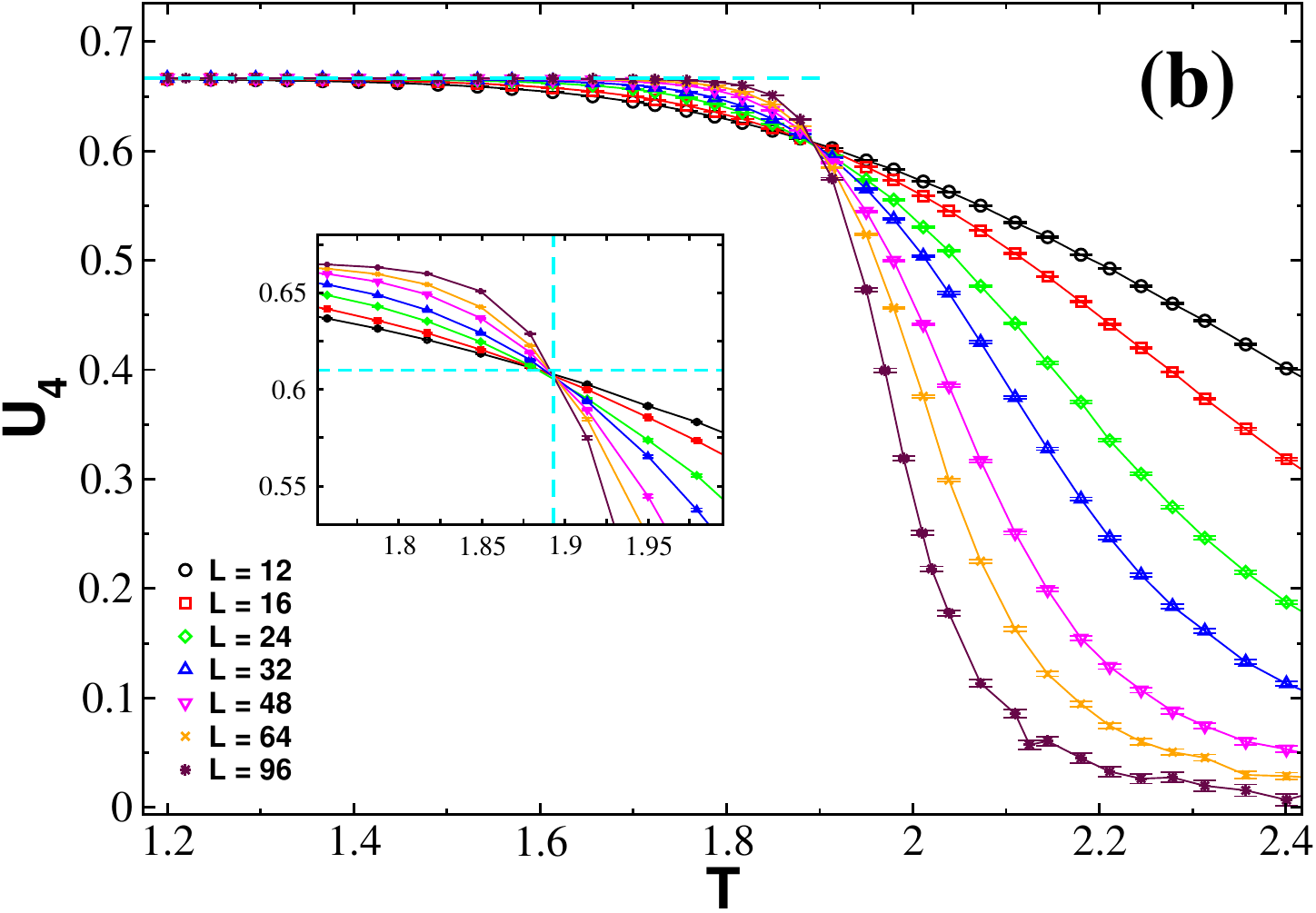}
	\caption{Plot of the Binder cumulant $U_4$ vs. temperature $T$ in disordered Baxter model with four-spin coupling parameter (a) $g_0 = 0.2$, and (b) $g_0 = -0.2$ for various systems sizes (see the keys) and disorder strength $\sqrt{u_0} = 0.2$. The vertical dashed line in main frames denotes the value $U_4(T\rightarrow 0, L\rightarrow \infty) = 2/3$. The inset in both panels magnifies the intersection region, where the estimated values of critical temperature $T_c$ and $U_* = U_4(T_c)$ are indicated by vertical and horizontal dashed lines, respectively.}
	\label{Fig_bind}
\end{figure}

Firstly, we determine the critical point $T_c$ by analyzing various observables of interest. In the Baxter model, the magnetization density ($m^{(1)}$ or $m^{(2)}$), defined on either of the lattices ($(1)$ or $(2)$), serves as an order parameter for the \textit{ferro--para} phase transition. An important observable for locating $T_c$ with good accuracy is the Binder cumulant $U_4$~\cite{Binder81}, defined as
\be
\label{bind_L}
U_4 = 1 - \frac{\overline{\left\langle m^4 \right\rangle}}{3 \overline{\left\langle m^2 \right\rangle}^2}
\; ,
\ee
where $\overline{(\cdots)}$ represents the disorder average and $\left\langle \cdots \right\rangle$ denotes the thermal average. Note that in the above expression $m$ is the magnetization density on either of the lattices. The ratio ${\overline{\left\langle m^4 \right\rangle}}/{\overline{\left\langle m^2 \right\rangle}^2}$ corresponds to the kurtosis of the order parameter distribution. In the ferromagnetic phase at low temperatures, the kurtosis approaches unity, and hence the Binder cumulant behaves as $U_4 \rightarrow 2/3$. At high temperatures $T$, the kurtosis approaches the Gaussian value $3$, i.e., $U_4 \rightarrow 0$. The Binder cumulant~\eqref{bind_L} is invariant under RG transformations and attains a universal, $L$-independent value at the critical point, $U_* = U_4(T_c)$.

We now turn to the numerical results. We consider two cases: $g_0 = 0.2$ and $g_0 = -0.2$, with a small disorder strength $\sqrt{u_0} = 0.2$, so that the analytical limits $\vert g_0 \vert \ll 1$ and $u_0 \ll 1$ remain applicable. In Fig.~\ref{Fig_bind}(a), the Binder cumulant $U_4$ is plotted as a function of temperature $T$ for $g_0 = 0.2$. Different datasets correspond to different system sizes. We observe that at sufficiently low temperatures, $U_4$ equals its limiting value $2/3$, independently of $L$, while at high temperatures it rapidly drops to zero due to the Gaussian nature of the order parameter. The critical temperature $T_c$ delimiting these two regimes is determined from the common intersection point of the different datasets, yielding $T_c \simeq 2.551(2)$ and $U_* \simeq 0.616(3)$. This value of $T_c$ is slightly lower than the pure-limit value $T_c(u_0=0) \simeq 2.577$~\cite{Wu71,Baxter71}. The data near the intersection point are magnified in the inset. 

In Fig.~\ref{Fig_bind}(b), we present results for $g_0 = -0.2$, which give $T_c \simeq 1.893(6)$ and $U_* \simeq 0.610(4)$. For comparison, the pure-case value is $T_c(u_0 = 0) \simeq 1.932$. Note that changing the sign of $g_0$ mainly shifts the location of the critical temperature, while the overall structure of the Binder-cumulant  remains qualitatively unchanged, as compared to the case~(a). Additionally, we have also verified that the data for $U_4$ in Fig.~\ref{Fig_bind} satisfy the finite-size scaling form $U_4 = f_{U}[(T - T_c)L^{1/\nu}]$, with a critical exponent $\nu \simeq 1$ for both $g_0 = \pm 0.2$.

\begin{figure}[t!]
\centering
	\includegraphics[width=0.48\textwidth]{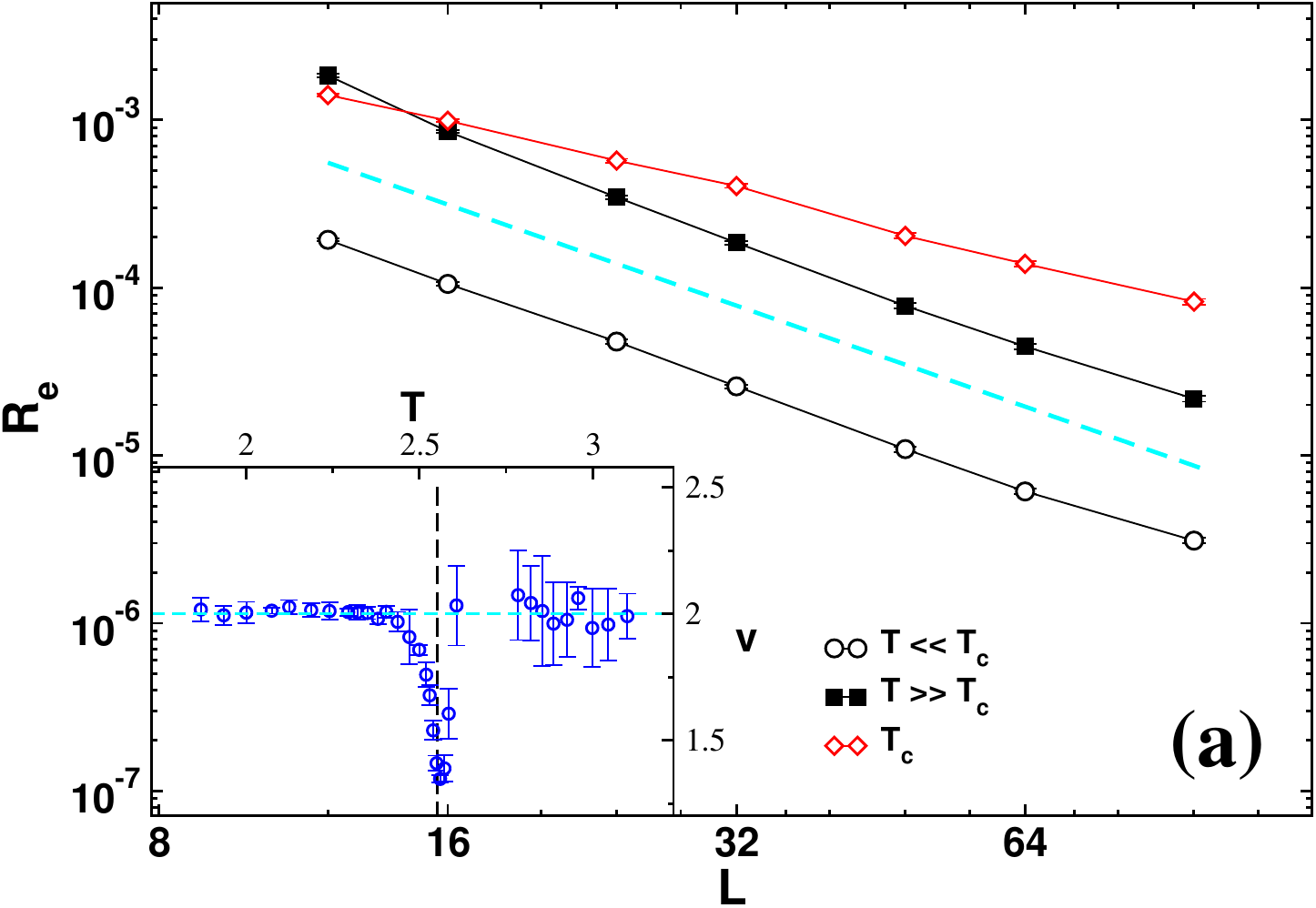}
	\includegraphics[width=0.48\textwidth]{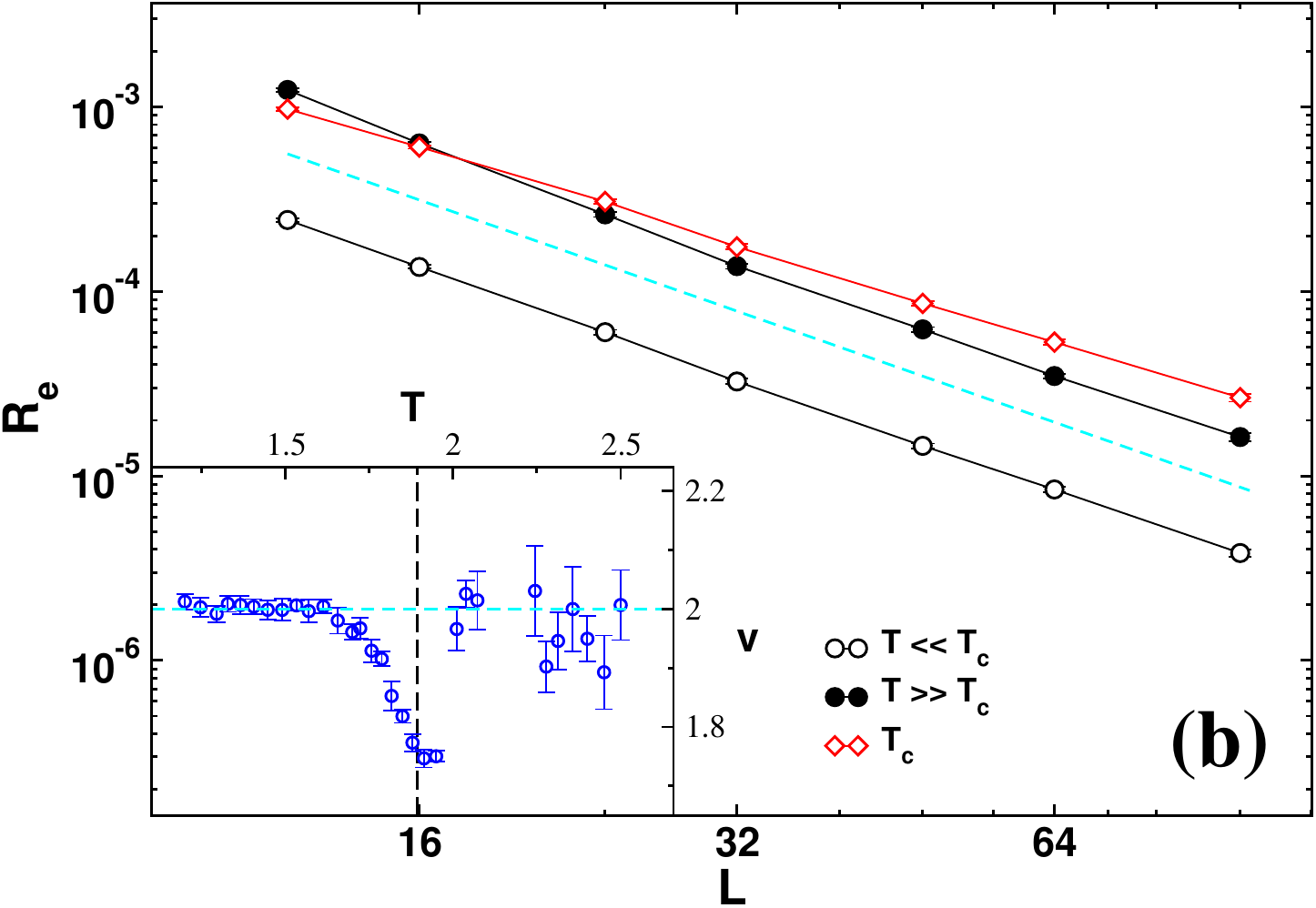}
	\caption{Relative variance $R_e$ of total internal energy vs. system size $L$ (on log-log scale) in disordered Baxter model with four-spin coupling strength (a) $g_0 = 0.2$, and (b) $g_0 = -0.2$ for different fixed temperatures (see the keys) and disorder strength $\sqrt{u_0} = 0.2$. The dashed line in main panels denotes the self-averaging law: $R_e \propto L^{-2}$. In the insets of both panels the power-law exponent $v$ obtained from a fit of form $R_e \propto L^{-v}$ is plotted against the temperature $T$. See main text for more details.}
	\label{Fig_relg0.2}
\end{figure}

We now present our main results. In Fig.~\ref{Fig_relg0.2}, the relative variance $R_e$ of the total internal energy $e$ (per volume) is shown. In the main frame of panel (a), this quantity is plotted as a function of system size $L$ for fixed parameters $g_0 = 0.2$, $\sqrt{u_0} = 0.2$, and various temperatures. At low $(T \ll T_c)$ and high $(T \gg T_c)$ temperatures, $R_e$ decreases as a power law with exponent $v = 2$ (the dimensionality $d$ of the system), indicating strong self-averaging behavior, as expected. Near $T_c$, $R_e$ still follows a power-law decay but with a smaller exponent $v \simeq 1.38$, suggesting that the total internal energy is \textit{weakly} self-averaging in the vicinity of the critical point.

In the inset of the same panel, we fit different datasets across $T_c$ to the power law $R_e \propto L^{-v}$ and plot the extracted exponent $v$ as a function of temperature $T$. During the fitting procedure, we ensure that the \textit{reduced}-$\chi^2$ values lie within an acceptable range. Interestingly, the exponent $v$ exhibits a pronounced dip near $T_c$ (indicated by a vertical dashed line). This behavior can be understood as follows. The total internal energy $e$ consists of regular and critical contributions, i.e., $e = e_0 + e_s$. The regular part $e_0$ is independent of $L$ and therefore remains self-averaging at all temperatures; moreover, it is analytic near $T_c$. In contrast, the critical part $e_s$ is singular at $T_c$ and contributes significantly only in its vicinity, as reflected in the inset. When $e_s$ is negligible, the total energy $e$ exhibits strong self-averaging with exponent $v \simeq d$. However, close to $T_c$, the exponent $v$ is reduced due to the competing effect of the critical contribution, indicating that it has a qualitatively different nature from the regular part. A similar behavior is observed for $g_0 = -0.2$, as shown in panel (b) of Fig.~\ref{Fig_relg0.2}.

\begin{figure}[t!]
\centering
	\includegraphics[width=0.48\textwidth]{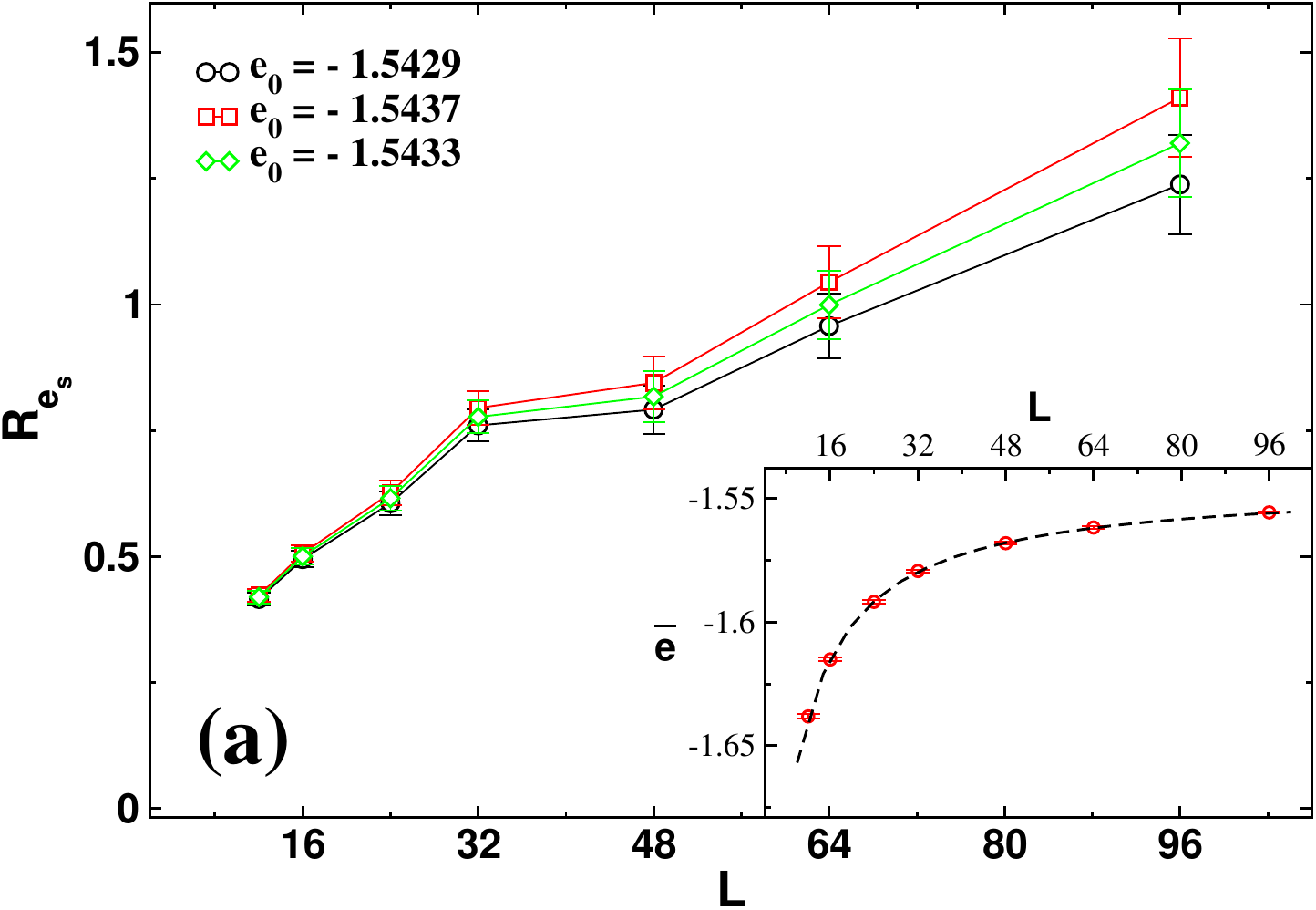}
	\includegraphics[width=0.48\textwidth]{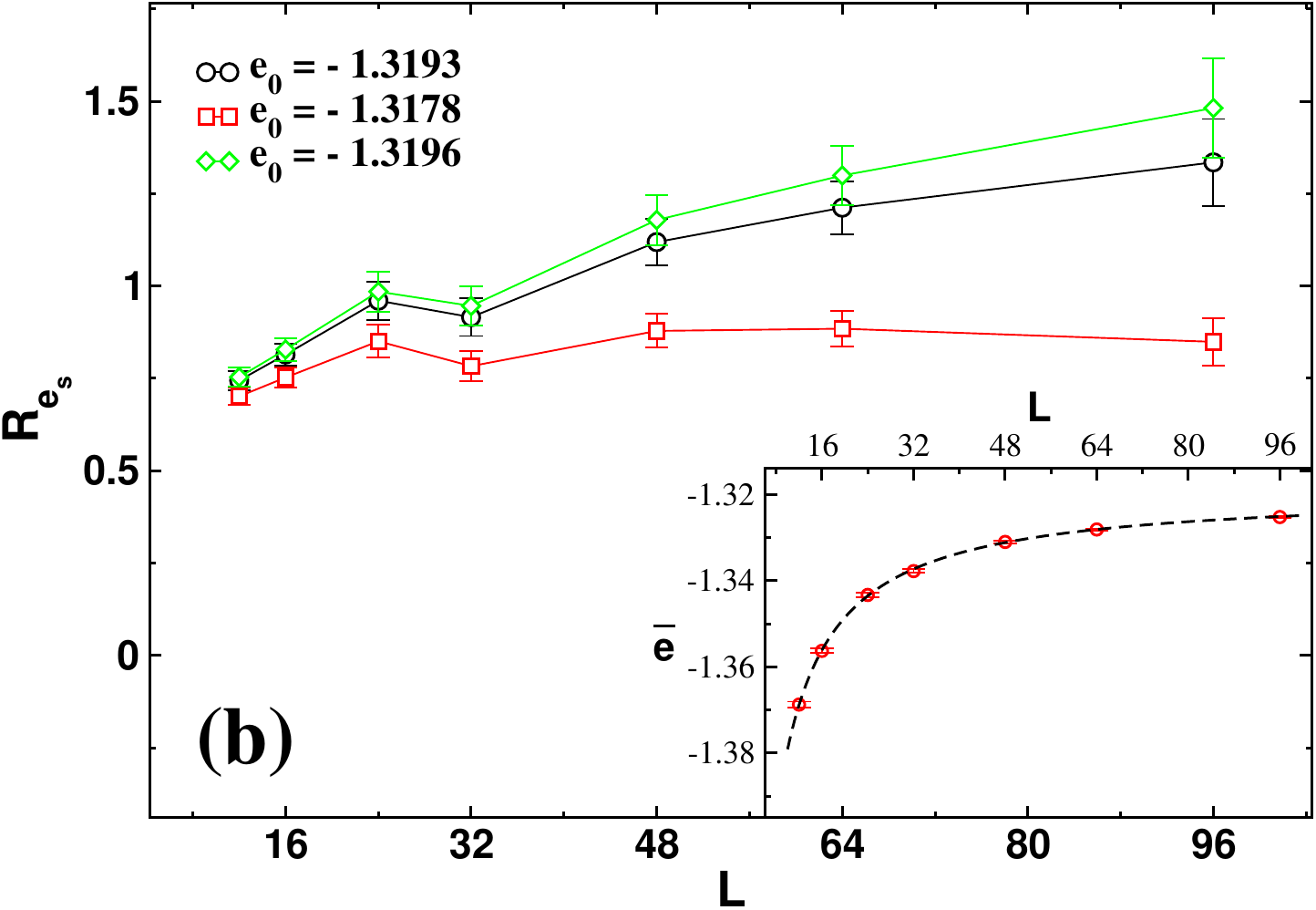}
	\caption{Relative variance $R_{e_s}$ of the critical internal energy vs. system size $L$ in disordered Baxter model with four-spin coupling (a) $g_0 = 0.2$, and (b) $g_0 = -0.2$ for different estimates of regular part $e_0$ (see the key) and fixed disorder strength $\sqrt{u_0} = 0.2$. Insets depict $\overline{ e }$ vs. $L$ at $T_c$. The dashed line denotes the fit of form~\eqref{fit_e}.}
	\label{Fig_g0.2fit}
\end{figure}

To probe the behavior of the critical energy $e_s$, it is necessary to separate the regular part $e_0$ from the total energy $e$. However, this is a challenging task, as the \textit{precise} numerical determination of $e_0$ is difficult. We therefore adopt the ansatz
\be
\overline{ e(L) } = e_0 + \alpha L^{-\Delta_\epsilon}
\; 
\label{fit_e}
\ee
for disordered average total energy $\overline{ e(L) }$. Here, $\Delta_\epsilon (= d - 1/\nu)$ is the scaling dimension of the energy operator. While this ansatz is not exact and logarithmic corrections may appear in the second term on the right-hand side, as observed in the Ising model (see Appendix~\ref{2DRIM}), such logarithmic corrections can be \textit{effectively} absorbed into the parameter $\Delta_\epsilon$ when fitting the data using the form above. We also remark that more elaborate ansatz including additional logarithmic corrections would introduce extra free parameters that cannot be reliably constrained by the present data.

\begin{table}[h!]
	\centering
	\caption{Fit of the internal energy to the form $\overline{e(L)} = e_0 + \alpha L^{-\Delta_\epsilon}$~\eqref{fit_e}. The upper half is for $g_0 = 0.2$ and the lower half is for $g_0 = -0.2$. The disorder strength $\sqrt{u_0}$ is fixed to $0.2$. During fits the parameter-significance $p$-value associated with $e_0$ remains near zero, corresponding to the null hypothesis $e_0 = 0$.}
	\vspace*{0.2cm}
	\begin{tabular}{| c | c | c | c | c | c | c | c | c |}
		\hline
		$L_{min}-L_{max}$   &   $e_0$      &  $\alpha$    & $\Delta_\epsilon$  & \textit{reduced}-$\chi^2$ \\ 
		\hline
		12-96    &    -1.5429 (6)   &  -1.06 (4) &  0.971 (16) & 0.23 \\
		12-48    &    -1.5437 (22)  &  -1.0 (1)  &  0.984 (45) & 0.43 \\
		24-96    &    -1.5433 (16)  &  -1.1 (2)  &  0.985 (66) & 0.44 \\
		\hline
		\hline
		12-96    &    -1.3193 (5)   &  -0.64 (4) &  1.033 (30) & 0.54 \\
		12-48    &    -1.3178 (18)  &  -0.57 (8) &  0.975 (71) & 0.74 \\
		24-96    &    -1.3196 (12)  &  -0.69 (23) & 1.059 (11) & 1.01 \\
		\hline
	\end{tabular}
	\label{FitE_Baxter}
\end{table}

In Table~\ref{FitE_Baxter} we present several fits over different ranges of $L$ for the parameters $g_0 = 0.2$ (upper panel) and $g_0 = -0.2$ (lower panel). The regular part $e_0$ varies significantly with the fitting range, and the scaling dimension $\Delta_\epsilon$ attains a value around $1$ for both $g_0 = \pm 0.2$ at $\sqrt{u_0} = 0.2$, with slight variations depending on the fitting range. Since we are limited to system sizes up to $L = 96$, it is difficult to conclude definitively whether the Ising universality class ($\Delta_\epsilon = 1$) is fully recovered, or the estimates are in a \textit{crossover} regime between the pure (Baxter) fixed point and the Ising one. When $g_0 = 0.2$, disorder is a relevant perturbation in the sense of the Harris criterion~\cite{Harris74} and the RG trajectories flow towards a new (Ising) fixed point. For $g_0 = -0.2$, disorder is rather an irrelevant perturbation and the Harris criterion requires that the critical behavior remain unchanged. Matthews et al.~\cite{swen} observed, for a special type of impurity, for $g_0 >0$ the Ising fixed point is indeed approached, but for $g_0 < 0$ the flow remains attracted towards Baxter’s pure fixed point, though the Ising fixed point (at $g_0 =0$) slows down the convergence. The smaller the magnitude of $g_0$ in the negative $g_0$ case, the slower is the convergence. On the contrary, in Ref.~\cite{DD84} it was argued that the RG flow approaches an Ising fixed point irrespective of the sign of $g_0$. We finally note that RG trajectories approach a \textit{true} fixed point only in the limit $L \rightarrow \infty$.

In Fig.~\ref{Fig_g0.2fit}, we plot the relative variance $R_{e_s}$ of the critical part $e_s = e - e_0$, obtained by subtracting the estimated regular part $e_0$ (see Table~\ref{FitE_Baxter}) from the total energy of each sample. Panel (a) shows results for $g_0 = 0.2$, while panel (b) shows results for $g_0 = -0.2$. We observe that small changes in the estimate of $e_0$ lead to significant variations in $R_{e_s}$. Importantly, in all cases presented, $R_{e_s}$ either increases or saturates with increasing $L$, indicating the non-self-averaging nature of the critical energy $e_s$.

\section{Summary and discussion}
\label{conclusion}

Critical phenomena in disordered spin systems have been attracting significant interest over the years, yet many of their key features remained poorly understood. One such feature is the \textit{self-averaging} behavior of thermodynamic observables. In this paper, we investigated the self-averaging properties of the internal energy in the weakly disordered Baxter eight-vertex model using both analytical and numerical approaches.

Using a combined framework based on the replica trick and renormalization group (RG), we first derived the RG flow equations in the double limit $|g_0| \ll 1$ and $|g_0| \ln L \gg 1$. These equations were solved then for small positive $g_0$, leading to explicit expressions for the first two moments of the critical internal energy. On this basis, we concluded that the relative variance of the critical energy \textit{asymptotically} approaches a constant, demonstrating that the critical energy is non-self-averaging in case of small positive $g_0$.

We also provided convincing numerical evidence for the non-self-averaging of the critical energy by simulating the disordered Baxter model on finite-size lattices, for both negative and positive values of the coupling parameter $g_0$. The total internal energy, composed of both regular and critical contributions, exhibits strong self-averaging at temperatures far from the critical point $T_c$, as expected from the central limit theorem. Near $T_c$, however, it displays a weak form of self-averaging due to the contribution of the singular part. To probe the behavior of the critical energy, we attempted to separate the singular component from the total energy by estimating the regular, $L$-independent part. Although \textit{naive} fits of the regular part remain approximate, they still allow the non-self-averaging behavior of the critical component to be observed. A more precise determination would require either the exact regular part, which is currently unavailable for the disordered model, or numerical data for much larger system sizes (e.g., $L \gtrsim 10^3$), where logarithmic corrections become negligible.

For completeness, we also numerically confirmed the non-self-averaging behavior of the critical energy in the two-dimensional disordered Ising model~\cite{2DRIM-IntEnergy}.

As a future research direction, it would be interesting to investigate the negative-$g_0$ regime of the disordered Baxter model in greater detail, in order to robustly determine the nature of the associated universality class.

\vspace{0.75cm}

\noindent
{\bf Acknowledgements.}
EM and RA acknowledge financial support from first FIS (Italian Science Fund) 2021 funding scheme (FIS783 - SMaC - Statistical Mechanics and Complexity) from MUR (Italian Ministry of University and Research).	MD acknowledges the support of the U.S.Department of Energy (DOE), Office of Science, Basic Energy Sciences, Materials Science and Engineering Division, under Award No. DE-SC0013599 (Subaward No. UTAUS SUB00000795).

\vspace{1cm}

\appendix
\renewcommand{\thesection}{\Alph{section}}
\section{Detailed analysis for the disordered Ising model}
\label{2DRIM}

In a reference~\cite{2DRIM-IntEnergy}, it was analytically argued that the disorder distribution of the critical internal energy, $P_{\tau}(E)$, is Gaussian in the two-dimensional disordered Ising model. Both the average critical energy and its sample-to-sample fluctuations were predicted to scale with the linear system size $L$ as $L \ln \ln (L)$, indicating that the critical internal energy is non-self-averaging.

In this section, we provide numerical evidence supporting these predictions. For this purpose, we perform extensive simulations of the random-bond Ising model on finite lattices. The nearest-neighbor bonds $J_{ij}$ are drawn from a bimodal distribution
\begin{equation}
	\label{dist_Ising}
	P(J_{ij}) = \frac{1}{2} \delta(J_{ij} - J_1) + \frac{1}{2} \delta(J_{ij} - J_2) \; ,
\end{equation}
where $J_1, J_2 > 0$, and the ratio $r = J_2 / J_1$ controls the strength of the disorder, with $r = 1$ corresponding to the pure case. The advantage of using the bimodal distribution~\eqref{dist_Ising} is that cluster Monte Carlo algorithms can be employed to suppress critical slowing down and access large system sizes. Additionally, the standard internal energy can be supplemented by the \textit{bond energy} $E_b$, defined as the number of bonds present in the Fortuin-Kasteleyn representation, which provides an alternative measure of the internal energy. Unlike the disordered Baxter model, the regular part of the bond energy in the disordered Ising model is exactly known ($E_b^{\rm reg} = L^2$). Therefore, the critical contribution, $E_b - L^2$, can be directly computed from the simulations. Furthermore, due to self-duality, the critical temperature $T_c$ is exactly given by~\cite{kinzel}
\be
\label{dual_ising}
({\rm e}^{J_1/T_c} - 1)({\rm e}^{r J_1/T_c} - 1) = 2
\; ,
\ee
which eliminates the need for a separate numerical determination of $T_c$.

\begin{figure}[t!]
	\begin{center}
		\includegraphics[width=8cm,height=6cm]{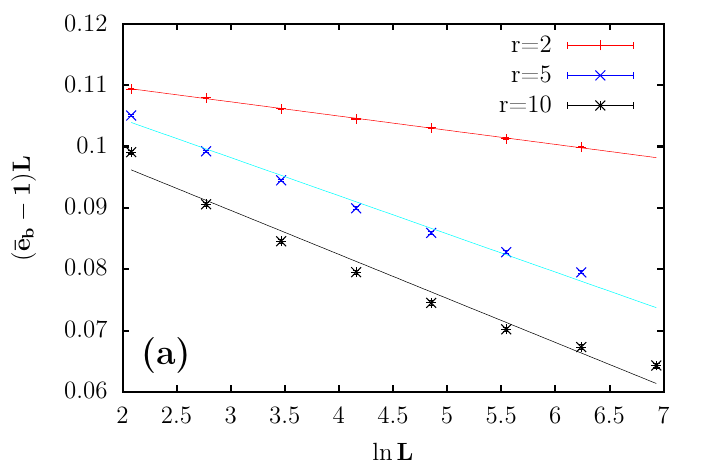}
		\includegraphics[width=8cm,height=6cm]{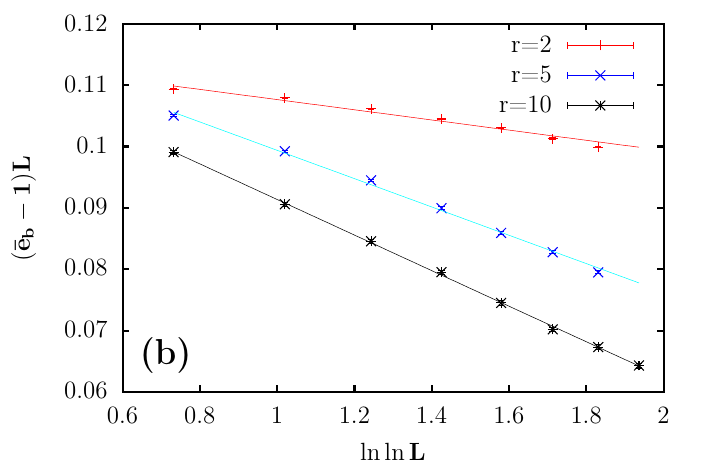}
	\end{center}
	\caption{Fit of disordered average bong energy in disordered Ising model: (a) plots $(\overline{ e_b } -1) L$ vs. $\ln {L}$, while (b) plots $(\overline{ e_b } -1) L$ vs. $\ln {\ln {L}}$. See text for details.}
	\label{ERBIM}
\end{figure}
\begin{figure}[t!]
	\begin{center}
		\includegraphics[width=8cm,height=6cm]{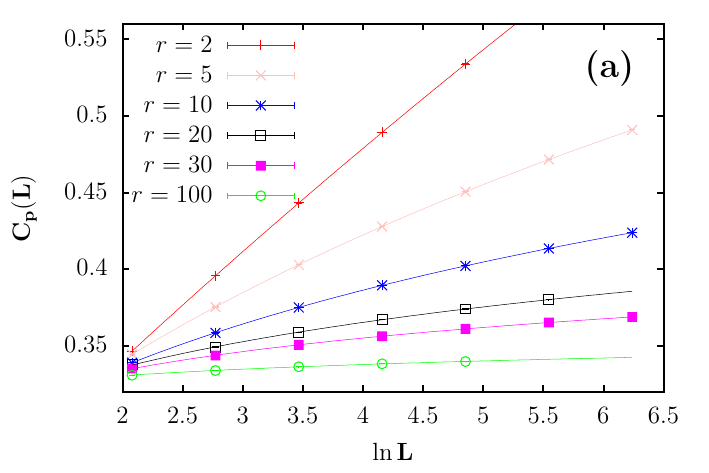}
		\includegraphics[width=8cm,height=6cm]{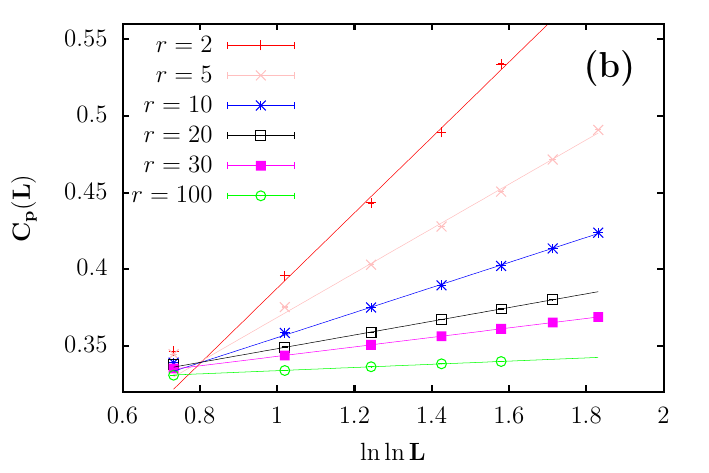}
	\end{center}
	\caption{Fit of specific heat $C_p(L)$ in disordered Ising model: (a) plots $C_p(L)$ vs. $\ln {L}$, while (b) plots $C_p(L)$ vs. $\ln {\ln {L}}$. See text for details.}
	\label{CSRBIM}
\end{figure}

We will separately consider the behavior of the disordered average energy and the sample-to-sample fluctuation of the energy, which, respectively, correspond to the denominator and the numerator of the relative variance $R_E$, see Eq.~\eqref{relVar}. Using the Swendsen-Wang algorithm~\cite{Swendsen}, we access systems of sizes up to $L=512$ for various values of disorder strength $r$, with around $10^7$ disorder realizations. We compare the measured values of the energy with analytical predictions~\cite{2DRIM-IntEnergy}. In the Ising model, since disorder is a \textit{marginal} perturbation, it is expected that the dimension of the energy is $\Delta_\epsilon=1$ and one needs to include logarithmic corrections. The disordered average critical energy (per volume) is predicted to be~\cite{2DRIM-IntEnergy}
\be
	\overline{ e_s (L) } = - \frac{1}{4 g_0 L} \ln{(1+ \frac{2}{\pi} g_0 \ln{L})} \; .
\ee
We consider in the following a fit of the average bond energy per spin $\overline{ e_b } = \overline{ E_b }/L^2$ (we obtain similar results by taking the standard internal energy, see below). This quantity goes to one in the large $L$ limit, so we will consider a fit to the form
\be
	\label{fitall}
	(\overline{ e_b }-1) L = a + b \ln{ (1 + c \ln{L}) } \; .
\ee
If $g_0 \simeq c$ is small, this behaves as
\be
	\label{fitl}
	(\overline{ e_b }-1) L = a + b c \ln{L} \; ,
\ee 
while for large $c$
\be
	\label{fitll}
	(\overline{ e_b }-1) L= a + b \ln{c} + b \ln{ \ln{L}} \; .
\ee
We first consider the case $r=2$, a small amount of disorder, and we obtain
\be
	(\overline{ e_b } -1) L  =  0.1683 (9) -0.24 (52) \ln{( 1 + 0.015 (35) \ln{L})}   \; ,
\ee
where $c$ is small (and not well determined). Then, it make sense to consider the form Eq.~\eqref{fitl}, for which we obtain
\be
	(\overline{ e_b } -1) L  = 0.1679 (3)   -0.00340 (6) \ln{L} \; .
\ee
For this small value of disorder, it behaves as a logarithmic term. 

For $r=5$, we obtain 
\be
	(\overline{ e_b } -1) L = 0.132 (2)  -0.035 (2) \ln{(1 + 0.56 (10) \ln{L})} \; .
\ee

For $r=10$, we obtain
\be
	(\overline{ e_b } -1) L = 0.11 (2) - 0.021 (2) \ln{(1+ 5.1 (6.0) \ln{L})}  \; ,
\ee
for which $c$ is large. It is then better to consider a fit to the form Eq.~\eqref{fitll}. We obtain
\be
	\label{eb_func}
	(\overline{ e_b } -1) L  = 0.0830 (3)  -0.0200 (2) \ln{\ln{L}} \; ,
\ee
which corresponds to a log-log behavior. All these results are illustrated in Fig.~\ref{ERBIM}. In the panel (a), we show $(\overline{ e_b } -1) L$ vs. $\ln{L}$, with best fits to the form $a + b \ln{ L }$ and in the panel (b) it is vs. $\ln{\ln{L}}$ with best fits to the form $a + b \ln{ \ln{L}}$. We observe that $(\overline{ e_b } -1) L$ is proportional to $\ln{L}$ for $r=2$ and to $\ln{\ln{L}}$ for $r=10$ in agreement with the previous fits.

Next, we repeat the same fit for the standard internal energy per spin for $r=10$. We consider a fit to the form
\be
	\label{log_form}
	\overline{ e (L)} = e_0  +  \frac{a + b \ln{(1 + c \ln{L})}}{L}
\ee
which contains also $e_0$, the regular part of the internal energy.  A fit to this form gives (not shown here for brevity)
\begin{eqnarray}
	\nonumber
	\overline{ e (L)} &=& 1.425990 (2) +  \frac{0.736 (57)  - 0.137 (6) \ln{(1 + 4.2 (2.4) \ln{L})}}{L} \; ,
	\\
	\nonumber
	\\
	&=& 1.425990 (2) +  \frac{0.518 (1)  - 0.127 (1) \ln{\ln{L}}}{L} \; .
\end{eqnarray}
The critical part are the same (up to a multiplicative constant $\simeq 6.2$) as the one obtained for bond energy $\overline{ e_b }$.

We also consider the specific heat. For this we obtain data up to $L=512$ with $10^5$ disorder realizations. Without disorder, it behaves as $\ln{L}$, whereas with disorder it is predicted to behave as $\ln{\ln{L}}$~\cite{DD82,2DRIM-IntEnergy,Wang1990}. We present the specific heat $C_p(L)$ as a function of $\ln{L}$ in panel (a) and as a function of $\ln{\ln{L}}$ in panel (b) of Fig.~\ref{CSRBIM} for various values of disorder $r$.

In Fig.~\ref{CSRBIM}(a), the continuous lines correspond to a best fit of the form 
\be
	C_p(L)  =  a + b \ln{(1 + c \ln{L})} \; .
\ee
We always obtain a good fit, with $c=0.056, 0.29,  0.69, 1.82, 10.3, 191$ for $r=2, 5, 10, 20, 30, 100$, respectively. One expects that $c = 1/\ln(L_c)$ with $L_c$ a crossover length. In Fig.~\ref{CSRBIM}(b), we show a fit to the form
\be
	C_p(L)  =  a + b \ln{\ln{L}} \; .
\ee
From this figure, we conclude that the log-log behavior is observed only for $r \simeq 10$ or larger, which is in agreement with the results from the energy and with the value $L_c \simeq 5$ for $r=10$.

We consider now the sample-to-sample fluctuation of the energy. In the large $L$-limit, the critical internal energy of a disorder realization can be written as
\begin{equation}
	\label{eng_cr}
	\mathcal{E}(L) \sim - \frac{1}{g_0} L \, \ln \ln (L) \, + \, L \, \ln \ln (L) \, f  \, = \, L^2 \overline{ e_s } + \delta \mathcal{E} \; ,
\end{equation}
with $f$ being a random variable from Gaussian distribution with variance unity. We have determined earlier that the form of first part (average critical energy) is in fact $a + b \ln{(1 + c \ln{L})}$. We now consider only the second part $\delta \mathcal{E}$, for which we compute
\be
	F_\mathcal{E}(L) = \overline{ \mathcal{E}^2 } - \overline{ \mathcal{E} }^2  = \overline{ \delta \mathcal{E}^2 }  = L^2 ( a + b \ln{(1 + c \ln{L})}) \; .
\ee
$F_\mathcal{E}(L)$ is the numerator of $R_E(L)$ as defined in Eq.\eqref{relVar}. We consider the case with disorder $r=10$ for the critical part of the bond energy, $e_b-1$. We obtain an excellent fit with 
\be
	\label{E_func}
	F_{L(e_b-1)}(L) = 0.2432 (7)  + 0.536 (3) \ln{(1 + 0.217 (2) \ln{L})}  \; .
\ee
From the results in Ref.~\cite{2DRIM-IntEnergy} one would expect that $F_\mathcal{E}(L)$ behaves as $(\ln{\ln{L}})^2$. But in fact, there is also a constant part, so we obtain $(\alpha + \beta \ln{\ln{L}})^2 \simeq \alpha^2 + 2 \alpha \beta  \ln{\ln{L}} + \cdots$. If $\beta$ is small (compared to $\alpha$) we will not observe the square log-log.

\begin{figure}[t!]
	\begin{center}
		\includegraphics[width=0.68\textwidth]{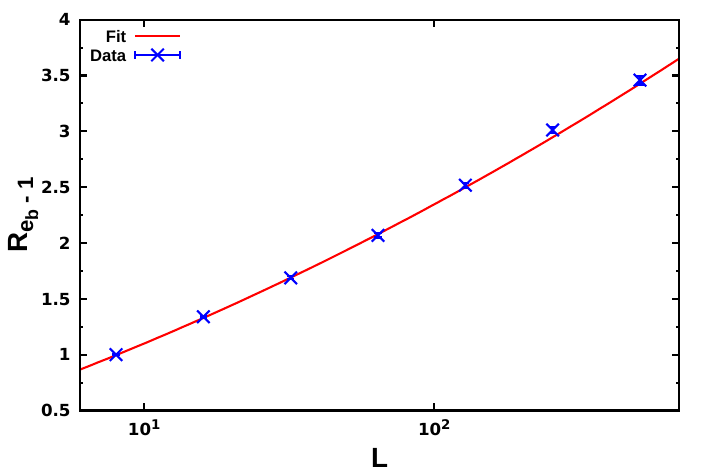}
	\end{center}
	\caption{Plot of $R_{e_b-1}$ vs. $L$ on a log-linear scale for the disordered Ising model with $r=10$. The line denotes the fit function $\frac{F_{L(e_b-1)}(L)}{((\overline{e_b} -1) L)^2}$. See text for details.}
	\label{SARe}
\end{figure}

For a clean demonstration of the self-averaging features, we further consider the relative variance~\eqref{relVar} of critical bond energy $e_b(L) - 1$. In Fig.~\ref{SARe}, this quantity is plotted against the system size $L$ on a log-linear scale for disorder strength $r=10$. We observe that $R_{e_b-1}$ is slightly increasing with $L$, thus critical bond energy is not self-averaging. We also show the plot of the fit function $\frac{F_{L(e_b-1)}(L)}{((\overline{ e_b } -1) L)^2}$ (where $F_{L(e_b-1)}(L)$ is given by Eq.~\eqref{E_func} and $(\overline{ e_b } -1) L$ by Eq.~\eqref{eb_func}) and compare to the data of $R_{e_b-1}$ (up to some constant $95.067247$). The agreement is perfect.

\addcontentsline{toc}{section}{References}
\bibliographystyle{iopart-num}


\begin{thebibliography}{10}
	\expandafter\ifx\csname url\endcsname\relax
	\def\url#1{{\tt #1}}\fi
	\expandafter\ifx\csname urlprefix\endcsname\relax\def\urlprefix{URL }\fi
	\providecommand{\eprint}[2][]{\url{#2}}

\bibitem{Harris74}
Harris A B 1974 {\em J. Phys. C: Solid State Phys.\/} {\bf 7} 1671
\urlprefix\url{https://doi.org/10.1088/0022-3719/7/9/009}

\bibitem{Harris_Lub74}
Harris A B and Lubensky T C 1974 {\em Phys. Rev. Lett.\/} {\bf 33} 1540--1543
\urlprefix\url{https://link.aps.org/doi/10.1103/PhysRevLett.33.1540}

\bibitem{Khmelnitskii75}
Khmel'nitskii D E 1975 {\em Sov. Phys. JETP\/} {\bf 41} 981

\bibitem{Lubensky75}
Lubensky T C 1975 {\em Phys. Rev. B\/} {\bf 11} 3573--3580
\urlprefix\url{https://link.aps.org/doi/10.1103/PhysRevB.11.3573}

\bibitem{Grinstein_Luther76}
Grinstein G and Luther A 1976 {\em Phys. Rev. B\/} {\bf 13} 1329--1343
\urlprefix\url{https://link.aps.org/doi/10.1103/PhysRevB.13.1329}

\bibitem{kinzel}
Kinzel W and Domany E 1981 {\em Phys. Rev. B\/} {\bf 23} 3421--3434
\urlprefix\url{https://link.aps.org/doi/10.1103/PhysRevB.23.3421}

\bibitem{DD82}
Dotsenko Vik. S and Dotsenko Vl. S 1982 {\em J. Phys. C: Solid State Phys.\/} {\bf 15} 495
\urlprefix\url{https://doi.org/10.1088/0022-3719/15/3/015}

\bibitem{Newman_Riedel82}
Newman K E and Riedel E K 1982 {\em Phys. Rev. B\/} {\bf 25} 264--280
\urlprefix\url{https://link.aps.org/doi/10.1103/PhysRevB.25.264}

\bibitem{Birgeneau83}
Birgeneau R J, Cowley R A, Shirane G, Yoshizawa H, Belanger D P, King A R and Jaccarino V 1983 {\em Phys. Rev. B\/} {\bf 27} 6747--6753
\urlprefix\url{https://link.aps.org/doi/10.1103/PhysRevB.27.6747}

\bibitem{rep1}
Pelissetto A and Vicari E 2002 {\em Phys. Rep.\/} {\bf 368} 549--727
\urlprefix\url{https://doi.org/10.1016/S0370-1573(02)00219-3}


\bibitem{perturbedCFT1}
Bernard D 1997 in {\em Low-Dimensional Applications of Quantum Field Theory},
NATO ASI Series, Series B: Physics Vol. 469,
ed L Baulieu, V Kazakov, M Picco and P Windey
(Boston: Springer) pp 19--61
\urlprefix\url{https://doi.org/10.1007/978-1-4899-1919-9_2}

\bibitem{perturbedCFT2}
Delfino G 2017 {\em Phys. Rev. Lett.\/} {\bf 118} 250601
\urlprefix\url{https://link.aps.org/doi/10.1103/PhysRevLett.118.250601}

\bibitem{IGLOI2005277}
Igl\'oi F and Monthus C 2005 {\em Phys. Rep.\/} {\bf 412} 277--431
\urlprefix\url{https://doi.org/10.1016/j.physrep.2005.02.006}

\bibitem{Picco_2006}
Picco M, Honecker A and Pujol P 2006 {\em J. Stat. Mech.: Theory Exp.\/} {\bf 2006} P09006
\urlprefix\url{https://doi.org/10.1088/1742-5468/2006/09/P09006}

\bibitem{PhysRevE.108.064131}
Agrawal R, Cugliandolo L F, Faoro L, Ioffe L B and Picco M 2023 {\em Phys. Rev. E\/} {\bf 108} 064131
\urlprefix\url{https://link.aps.org/doi/10.1103/PhysRevE.108.064131}

\bibitem{NonSelfAverage1}
Wiseman S and Domany E 1995 {\em Phys. Rev. E\/} {\bf 52} 3469--3484
\urlprefix\url{https://link.aps.org/doi/10.1103/PhysRevE.52.3469}

\bibitem{NonSelfAverage2}
Aharony A and Harris A B 1996 {\em Phys. Rev. Lett.\/} {\bf 77} 3700--3703
\urlprefix\url{https://link.aps.org/doi/10.1103/PhysRevLett.77.3700}

\bibitem{NonSelfAverage3}
Wiseman S and Domany E 1998 {\em Phys. Rev. Lett.\/} {\bf 81} 22--25
\urlprefix\url{https://link.aps.org/doi/10.1103/PhysRevLett.81.22}

\bibitem{NonSelfAverage4}
Dillmann O, Janke W and Binder K 1998 {\em J. Stat. Phys.\/} {\bf 92} 57--100
\urlprefix\url{https://doi.org/10.1023/A:1023043602398}

\bibitem{Dotsenko_Klumov2012}
Dotsenko V and Klumov B 2012 {\em J. Stat. Mech.: Theory Exp.\/} {\bf 2012} P05027
\urlprefix\url{https://doi.org/10.1088/1742-5468/2012/05/P05027}

\bibitem{Dotsenko_Holovatch14}
Dotsenko V and Holovatch Y 2014 {\em Phys. Rev. E\/} {\bf 90} 052126
\urlprefix\url{https://link.aps.org/doi/10.1103/PhysRevE.90.052126}

\bibitem{2DRIM-IntEnergy}
Dotsenko V, Holovatch Y, Dudka M and Weigel M 2017 {\em Phys. Rev. E\/} {\bf 95} 032118
\urlprefix\url{https://link.aps.org/doi/10.1103/PhysRevE.95.032118}

\bibitem{Onsager44}
Onsager L 1944 {\em Phys. Rev.\/} {\bf 65} 117--149
\urlprefix\url{https://link.aps.org/doi/10.1103/PhysRev.65.117}

\bibitem{DD_rev83}
Dotsenko Vik. S and Dotsenko Vl. S 1983 {\em Adv. Phys.\/} {\bf 32} 129--172
\urlprefix\url{https://doi.org/10.1080/00018738300101541}

\bibitem{2DRIM-rev1}
Hasenbusch M, Toldin F P, Pelissetto A and Vicari E 2008 {\em Phys. Rev. E\/} {\bf 78} 011110
\urlprefix\url{https://link.aps.org/doi/10.1103/PhysRevE.78.011110}

\bibitem{2DRIM-rev2}
Kenna R and Ruiz-Lorenzo J J 2008 {\em Phys. Rev. E\/} {\bf 78} 031134
\urlprefix\url{https://link.aps.org/doi/10.1103/PhysRevE.78.031134}

\bibitem{correlated1}
Bagam\'ery F A, Turban L and Igl\'oi F 2005 {\em Phys. Rev. B\/} {\bf 72} 094202
\urlprefix\url{https://link.aps.org/doi/10.1103/PhysRevB.72.094202}

\bibitem{correlated2}
Dudka M, Fedorenko A A, Blavatska V and Holovatch Y 2016 {\em Phys. Rev. B\/} {\bf 93} 224422
\urlprefix\url{https://link.aps.org/doi/10.1103/PhysRevB.93.224422}

\bibitem{correlated3}
Chippari F, Picco M and Santachiara R 2023 {\em SciPost Phys.\/} {\bf 15} 135
\urlprefix\url{https://doi.org/10.21468/SciPostPhys.15.4.135}

\bibitem{fed}
Dudka M and Fedorenko A A 2017 {\em Condens. Matter Phys.\/} {\bf 20} 13603
\urlprefix\url{https://doi.org/10.5488/CMP.20.13603}

\bibitem{brout}
Brout R 1959 {\em Phys. Rev.\/} {\bf 115} 824--835
\urlprefix\url{https://link.aps.org/doi/10.1103/PhysRev.115.824}

\bibitem{Baxter71}
Baxter R J 1971 {\em Phys. Rev. Lett.\/} {\bf 26} 832--833
\urlprefix\url{https://link.aps.org/doi/10.1103/PhysRevLett.26.832}

\bibitem{Kadanoff_Wegner71}
Kadanoff L P and Wegner F J 1971 {\em Phys. Rev. B\/} {\bf 4} 3989--3993
\urlprefix\url{https://link.aps.org/doi/10.1103/PhysRevB.4.3989}

\bibitem{Wu71}
Wu F Y 1971 {\em Phys. Rev. B\/} {\bf 4} 2312--2314
\urlprefix\url{https://link.aps.org/doi/10.1103/PhysRevB.4.2312}

\bibitem{swen}
Matthews-Morgan D, Landau D P and Swendsen R H 1984 {\em Phys. Rev. Lett.\/} {\bf 53} 679--682
\urlprefix\url{https://link.aps.org/doi/10.1103/PhysRevLett.53.679}

\bibitem{Luther_Peschel75}
Luther A and Peschel I 1975 {\em Phys. Rev. B\/} {\bf 12} 3908--3917
\urlprefix\url{https://link.aps.org/doi/10.1103/PhysRevB.12.3908}

\bibitem{Luther76}
Luther A 1976 {\em Phys. Rev. B\/} {\bf 14} 2153--2159
\urlprefix\url{https://link.aps.org/doi/10.1103/PhysRevB.14.2153}

\bibitem{DD84}
Dotsenko Vik. S and Dotsenko Vl. S 1984 {\em J. Phys. A: Math. Gen.\/} {\bf 17} L301
\urlprefix\url{https://doi.org/10.1088/0305-4470/17/5/016}

\bibitem{Larkin-Khmelnitskii}
Larkin A I and Khmelnitskii D E 1969 {\em Sov. Phys. JETP\/} {\bf 29} 1123

\bibitem{Aharony}
Aharony A 1976 {\em Phys. Rev. B\/} {\bf 13} 2092--2098
\urlprefix\url{https://link.aps.org/doi/10.1103/PhysRevB.13.2092}

\bibitem{Hukushima96}
Hukushima K and Nemoto K 1996 {\em J. Phys. Soc. Jpn.\/} {\bf 65} 1604--1608
\urlprefix\url{https://doi.org/10.1143/JPSJ.65.1604}

\bibitem{Marinari92}
Marinari E and Parisi G 1992 {\em Europhys. Lett.\/} {\bf 19} 451
\urlprefix\url{https://doi.org/10.1209/0295-5075/19/6/002}

\bibitem{efron1982jackknife}
Efron B 1982 {\em The Jackknife, the Bootstrap and Other Resampling Plans\/} SIAM, Philadelphia

\bibitem{Binder81}
Binder K 1981 {\em Phys. Rev. Lett.\/} {\bf 47} 693--696
\urlprefix\url{https://link.aps.org/doi/10.1103/PhysRevLett.47.693}

\bibitem{Swendsen}
Swendsen R H and Wang J S 1987 {\em Phys. Rev. Lett.\/} {\bf 58} 86--88
\urlprefix\url{https://link.aps.org/doi/10.1103/PhysRevLett.58.86}

\bibitem{Wang1990}
Wang J S, Selke W, Dotsenko V S and Andreichenko V B 1990 {\em Physica A\/} {\bf 164} 221--239
\urlprefix\url{https://doi.org/10.1016/0378-4371(90)90196-Y}

\end{thebibliography}
\providecommand{\newblock}{}

\end{document}